%% file: main.tex
\documentclass[12pt,a4paper]{article}

\usepackage{ifthen} 
\newboolean{pdflatex}
\setboolean{pdflatex}{true} 

\newboolean{articletitles}
\setboolean{articletitles}{true} 

\newboolean{uprightparticles}
\setboolean{uprightparticles}{false} 

\def\paperauthors{LHCb collaboration} 
\def\paperasciititle{Observation of a new excited charm-strange meson D_{s1}(2933)^+ \\ in  B0->D+D-K+pi- decays} 
\def\papertitle{Observation of a new excited charm-strange meson $D_{s1}(2933)^+$ \\ in  $B^0\to D^+ D^- K^+ \pi^-$ decays} 
\def\paperkeywords{{High Energy Physics}, {LHCb}} 
\def\papercopyright{\the\year\ CERN for the benefit of the LHCb collaboration} 
\def\paperlicence{CC BY 4.0 licence}
\def\paperlicenceurl{https://creativecommons.org/licenses/by/4.0/}

\newif\ifEnableSectionTOCLinks
\EnableSectionTOCLinksfalse 

\input{LHCb/preamble}
\input{my-symbols-def}

\usepackage{longtable} 
\usepackage{rotating}
\usepackage{datatool}
\usepackage{multirow}
\usepackage{booktabs}

\DTLsetseparator{,}
\DTLloaddb{config}{results.csv}
\newcommand{\val}[1]{\DTLfetch{config}{key}{#1}{value}}

\begin{document}
\begin{filecontents*}{results.csv}
key,value
version_of_this_file,20251105
newDs_Mass_mev,2933
newDs_Mass_syst_upper_mev,4
newDs_Mass_syst_lower_mev,3
newDs_Mass_stat_upper_mev,6
newDs_Mass_stat_lower_mev,5
newDs_Width_mev,72
newDs_Width_syst_upper_mev,7
newDs_Width_syst_lower_mev,10
newDs_Width_stat_upper_mev,18
newDs_Width_stat_lower_mev,12
Ds2590_pole_Mass_mev,2606
Ds2590_pole_Mass_syst_upper_mev,11
Ds2590_pole_Mass_syst_lower_mev,11
Ds2590_pole_Mass_stat_upper_mev,2
Ds2590_pole_Mass_stat_lower_mev,5
Ds2590_pole_Width_mev,87
Ds2590_pole_Width_syst_upper_mev,5
Ds2590_pole_Width_syst_lower_mev,13
Ds2590_pole_Width_stat_upper_mev,5
Ds2590_pole_Width_stat_lower_mev,14
massfit_yield,3264
massfit_uncert,62
\end{filecontents*}

\renewcommand{\thefootnote}{\fnsymbol{footnote}}
\setcounter{footnote}{1}

\input{title-LHCb-PAPER}


\renewcommand{\thefootnote}{\arabic{footnote}}
\setcounter{footnote}{0}

\cleardoublepage


\pagestyle{plain} 
\setcounter{page}{1}
\pagenumbering{arabic}


\input{body}

\input{LHCb/acknowledgements}

\input{supplemental}

\addcontentsline{toc}{section}{References}
\bibliographystyle{LHCb/LHCb}
\bibliography{main,LHCb/standard,LHCb/LHCb-PAPER,LHCb/LHCb-CONF,LHCb/LHCb-DP,LHCb/LHCb-TDR}

\newpage
\input{Authorship_LHCb-PAPER-2025-073}

\end{document}

%% file: LHCb/preamble.tex
\usepackage[top=1in, bottom=1.25in, left=1in, right=1in]{geometry}

\usepackage{microtype}
\usepackage{lineno}  
\usepackage{xspace} 
\usepackage{caption} 

\usepackage{graphicx}  
\usepackage{color}
\usepackage{colortbl}
\graphicspath{{./figs/}} 

\usepackage{amsmath} 
\usepackage{amssymb}
\usepackage{amsfonts}
\usepackage{upgreek} 

\newcommand*\patchAmsMathEnvironmentForLineno[1]{%
\expandafter\let\csname old#1\expandafter\endcsname\csname #1\endcsname
\expandafter\let\csname oldend#1\expandafter\endcsname\csname
end#1\endcsname
 \renewenvironment{#1}%
   {\linenomath\csname old#1\endcsname}%
   {\csname oldend#1\endcsname\endlinenomath}%
}
\newcommand*\patchBothAmsMathEnvironmentsForLineno[1]{%
  \patchAmsMathEnvironmentForLineno{#1}%
  \patchAmsMathEnvironmentForLineno{#1*}%
}
\AtBeginDocument{%
\patchBothAmsMathEnvironmentsForLineno{equation}%
\patchBothAmsMathEnvironmentsForLineno{align}%
\patchBothAmsMathEnvironmentsForLineno{flalign}%
\patchBothAmsMathEnvironmentsForLineno{alignat}%
\patchBothAmsMathEnvironmentsForLineno{gather}%
\patchBothAmsMathEnvironmentsForLineno{multline}%
\patchBothAmsMathEnvironmentsForLineno{eqnarray}%
}

\usepackage[pdftex,
            pdfauthor={\paperauthors},
            pdftitle={\paperasciititle},
            pdfkeywords={\paperkeywords}]{hyperref}
\usepackage{hyperxmp}
\hypersetup{
    pdfcopyright={Copyright (C) \papercopyright},
    pdflicenseurl={\paperlicenceurl}
}

\usepackage[colorinlistoftodos,textsize=scriptsize]{todonotes}

\usepackage[bottom,flushmargin,hang,multiple]{footmisc}

\usepackage[all]{hypcap} 

\input{LHCb/lhcb-symbols-def} 

\hypersetup{
  colorlinks   = true, 
  urlcolor     = blue, 
  linkcolor    = blue, 
  citecolor    = red   
}

\ifEnableSectionTOCLinks
    \usepackage[explicit]{titlesec} 
    
    \let\oldcontentsline\contentsline
    \renewcommand

    \titleformat{\section}{\normalfont\Large\bf}{\hyperlink{tocsection.\thesection}{{\thesection} \parbox[t]{\dimexpr\textwidth-1pc}{#1}}}{1pc}{}

    \titleformat{\subsection}{\normalfont\bf}{\hyperlink{tocsubsection.\thesubsection}{{\thesubsection} \parbox[t]{\dimexpr\textwidth-1pc}{#1}}}{1pc}{}

    \titleformat{name=\section,numberless}[display]{}{}{0pt}{\normalfont\Huge\bfseries #1}
\fi

\usepackage{cite} 
\usepackage{LHCb/mciteplus}

%% file: LHCb/lhcb-symbols-def.tex
\usepackage{xspace} 
\usepackage{upgreek}

\def\lhcb   {\mbox{LHCb}\xspace}

\def\MagUp {\mbox{\em Mag\kern -0.05em Up}\xspace}

\ifthenelse{\boolean{uprightparticles}}%
{

 \def\Pmu         {\ensuremath{\upmu}\xspace}

 \def\Ppi         {\ensuremath{\uppi}\xspace}

 \def\Pchi        {\ensuremath{\upchi}\xspace}                 
 \def\Ppsi        {\ensuremath{\uppsi}\xspace}

 \def\PDelta      {\ensuremath{\Delta}\xspace}                 
 \def\PXi         {\ensuremath{\Xi}\xspace}                 
 \def\PLambda     {\ensuremath{\Lambda}\xspace}                 
 \def\PSigma      {\ensuremath{\Sigma}\xspace}                 
 \def\POmega      {\ensuremath{\Omega}\xspace}                 
 \def\PUpsilon    {\ensuremath{\Upsilon}\xspace}
 \let\oldPi\Pi
 \def\PPi         {\ensuremath{\oldPi}\xspace}

 \def\PB      {\ensuremath{\mathrm{B}}\xspace}                 
 \def\PD      {\ensuremath{\mathrm{D}}\xspace}                 
 \def\PJ      {\ensuremath{\mathrm{J}}\xspace}                 
 \def\PK      {\ensuremath{\mathrm{K}}\xspace}                 
 \def\Pb      {\ensuremath{\mathrm{b}}\xspace}                 
 \def\Pc      {\ensuremath{\mathrm{c}}\xspace}                 
 \def\Pd      {\ensuremath{\mathrm{d}}\xspace}                 

 \def\Pp      {\ensuremath{\mathrm{p}}\xspace}                 

 \def\Ps      {\ensuremath{\mathrm{s}}\xspace}                 
                  
 \def\Pu      {\ensuremath{\mathrm{u}}\xspace}                 
 \def\thebaroffset{0.0em}
}
{

 \def\Pmu         {\ensuremath{\mu}\xspace}

 \def\Ppi         {\ensuremath{\pi}\xspace}

 \def\Pchi        {\ensuremath{\chi}\xspace}                 
 \def\Ppsi        {\ensuremath{\psi}\xspace}                 
                  
 \mathchardef\PDelta="7101
 \mathchardef\PXi="7104
 \mathchardef\PLambda="7103
 \mathchardef\PSigma="7106
 \mathchardef\POmega="710A
 \mathchardef\PUpsilon="7107
 \mathchardef\PPi="7105
 \def\PB      {\ensuremath{B}\xspace}                 
 \def\PD      {\ensuremath{D}\xspace}                 
 \def\PJ      {\ensuremath{J}\xspace}                 
 \def\PK      {\ensuremath{K}\xspace}                 
 \def\Pb      {\ensuremath{b}\xspace}                 
 \def\Pc      {\ensuremath{c}\xspace}                 
 \def\Pd      {\ensuremath{d}\xspace}                 

 \def\Pp      {\ensuremath{p}\xspace}                 

 \def\Ps      {\ensuremath{s}\xspace}                 
                  
 \def\Pu      {\ensuremath{u}\xspace}                 
 \def\thebaroffset{0.18em}
}
\newcommand{\offsetoverline}[2][\thebaroffset]{\kern #1\overline{\kern -#1 #2}}%

\makeatletter
\ifcase \@ptsize \relax
  \newcommand{\miniscule}{\@setfontsize\miniscule{4}{5}}
\or
  \newcommand{\miniscule}{\@setfontsize\miniscule{5}{6}}
\or
  \newcommand{\miniscule}{\@setfontsize\miniscule{5}{6}}
\fi
\makeatother

\DeclareRobustCommand{\optbar}[1]{\shortstack{{\miniscule (\rule[.5ex]{1.25em}{.18mm})}
  \\ [-.7ex] $#1$}}

\def\mup        {{\ensuremath{\Pmu^+}}\xspace}
\def\mun        {{\ensuremath{\Pmu^-}}\xspace} 

\def\uquark    {{\ensuremath{\Pu}}\xspace}

\def\dquark    {{\ensuremath{\Pd}}\xspace}

\def\squark    {{\ensuremath{\Ps}}\xspace}
\def\squarkbar {{\ensuremath{\overline \squark}}\xspace}

\def\cquark    {{\ensuremath{\Pc}}\xspace}
\def\cquarkbar {{\ensuremath{\overline \cquark}}\xspace}

\def\bquark    {{\ensuremath{\Pb}}\xspace}

\def\pion   {{\ensuremath{\Ppi}}\xspace}

\def\pim    {{\ensuremath{\pion^-}}\xspace}
\def\pipm   {{\ensuremath{\pion^\pm}}\xspace}

\def\kaon    {{\ensuremath{\PK}}\xspace}

\def\KorKbar {\kern \thebaroffset\optbar{\kern -\thebaroffset \PK}{}\xspace}

\def\Kp      {{\ensuremath{\kaon^+}}\xspace}

\def\Kmp     {{\ensuremath{\kaon^\mp}}\xspace}

\def\Kstarz  {{\ensuremath{\kaon^{*0}}}\xspace}

\def\Kstar   {{\ensuremath{\kaon^*}}\xspace}

\def\D       {{\ensuremath{\PD}}\xspace}

\def\DorDbar {\kern \thebaroffset\optbar{\kern -\thebaroffset \PD}\xspace}

\def\Dp      {{\ensuremath{\D^+}}\xspace}
\def\Dm      {{\ensuremath{\D^-}}\xspace}
\def\Dpm     {{\ensuremath{\D^\pm}}\xspace}

\def\DpDm    {\ensuremath{\Dp {\kern -0.16em \Dm}}\xspace}

\def\Dstarz  {{\ensuremath{\D^{*0}}}\xspace}

\def\Ds      {{\ensuremath{\D^+_\squark}}\xspace}
\def\Dsp     {{\ensuremath{\D^+_\squark}}\xspace}

\def\B       {{\ensuremath{\PB}}\xspace}

\def\BorBbar {\kern \thebaroffset\optbar{\kern -\thebaroffset \PB}\xspace}
\def\Bz      {{\ensuremath{\B^0}}\xspace}

\def\Bd      {{\ensuremath{\B^0}}\xspace}

\def\BdorBdbar {\kern \thebaroffset\optbar{\kern -\thebaroffset \Bd}\xspace}
\def\Bu      {{\ensuremath{\B^+}}\xspace}

\def\Bp      {{\ensuremath{\Bu}}\xspace}

\def\Bs      {{\ensuremath{\B^0_\squark}}\xspace}

\def\BsorBsbar {\kern \thebaroffset\optbar{\kern -\thebaroffset \Bs}\xspace}

\def\jpsi     {{\ensuremath{{\PJ\mskip -3mu/\mskip -2mu\Ppsi}}}\xspace}

\def\chiczero {{\ensuremath{\Pchi_{\cquark 0}}}\xspace}

\def\chictwo  {{\ensuremath{\Pchi_{\cquark 2}}}\xspace}

\def\Y#1S{\ensuremath{\PUpsilon{(#1S)}}\xspace}

\def\proton      {{\ensuremath{\Pp}}\xspace}

\def\LorLbar     {\kern \thebaroffset\optbar{\kern -\thebaroffset \PLambda}\xspace}

\newcommand{\decay}[2]{\ensuremath{\mathinner{#1\!\to #2}}\xspace}

\def\to                 {\ensuremath{\rightarrow}\xspace}

\def\AT#1     {\ensuremath{A_{\mathrm{T}}^{#1}}\xspace}           

\def\C#1      {\ensuremath{\mathcal{C}_{#1}}\xspace}                       
\def\Cp#1     {\ensuremath{\mathcal{C}_{#1}^{'}}\xspace}                    
\def\Ceff#1   {\ensuremath{\mathcal{C}_{#1}^{\mathrm{(eff)}}}\xspace}        
\def\Cpeff#1  {\ensuremath{\mathcal{C}_{#1}^{'\mathrm{(eff)}}}\xspace}       
\def\Ope#1    {\ensuremath{\mathcal{O}_{#1}}\xspace}                       
\def\Opep#1   {\ensuremath{\mathcal{O}_{#1}^{'}}\xspace}                    

\newcommand{\aunit}[1]{\ensuremath{\text{\,#1}}}       

\newcommand{\tev}{\aunit{Te\kern -0.1em V}\xspace}
\newcommand{\gev}{\aunit{Ge\kern -0.1em V}\xspace}
\newcommand{\mev}{\aunit{Me\kern -0.1em V}\xspace}
\newcommand{\kev}{\aunit{ke\kern -0.1em V}\xspace}
\newcommand{\ev}{\aunit{e\kern -0.1em V}\xspace}
 
\newcommand{\mevc}{\ensuremath{\aunit{Me\kern -0.1em V\!/}c}\xspace}
\newcommand{\gevc}{\ensuremath{\aunit{Ge\kern -0.1em V\!/}c}\xspace}
\newcommand{\mevcc}{\ensuremath{\aunit{Me\kern -0.1em V\!/}c^2}\xspace}
\newcommand{\gevcc}{\ensuremath{\aunit{Ge\kern -0.1em V\!/}c^2}\xspace}

\def\fb   {\ensuremath{\aunit{fb}}\xspace}
\def\invfb   {\ensuremath{\fb^{-1}}\xspace}

\newcommand{\stat}{\aunit{(stat)}\xspace}
\newcommand{\syst}{\aunit{(syst)}\xspace}

\newcommand{\chisq}{\ensuremath{\chi^2}\xspace}

\newcommand{\chisqip}{\ensuremath{\chi^2_{\text{IP}}}\xspace}

\def\deriv {\ensuremath{\mathrm{d}}}

\def\gsim{{~\raise.15em\hbox{$>$}\kern-.85em
          \lower.35em\hbox{$\sim$}~}\xspace}
\def\lsim{{~\raise.15em\hbox{$<$}\kern-.85em
          \lower.35em\hbox{$\sim$}~}\xspace}

\def\sPlot{\mbox{\em sPlot}\xspace}

\def\sqs   {\ensuremath{\protect\sqrt{s}}\xspace}

\def\pt         {\ensuremath{p_{\mathrm{T}}}\xspace}

\def\mrad{\aunit{mrad}\xspace}

\def\tell1  {TELL1\xspace}
\def\ukl1   {UKL1\xspace}

\newcommand{\lhcborcid}[1]{\href{https://orcid.org/#1}{\hspace*{0.1em}\raisebox{-0.45ex}{\includegraphics[width=1em]{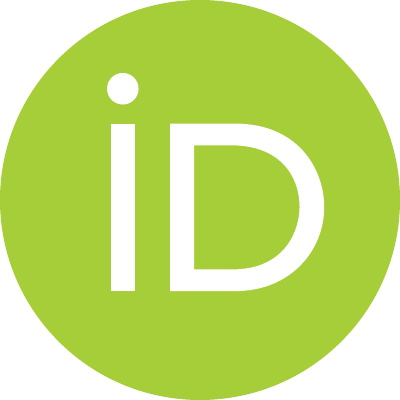}}}}


%% file: my-symbols-def.tex
\def\BdToDDKpi   {\decay{\Bd}{\Dp\Dm\Kp\pim}}
\newcommand{\mDKpi}{{\ensuremath{m(\Dp\Kp\pim)}\xspace}}

\newcommand{\DsMeson}[2]{\ensuremath{\D_{\squark#1}(#2)^+}\xspace}
\newcommand{\DsMesonNoMass}[1]{\ensuremath{\D_{\squark#1}^+}\xspace}
\newcommand{\DsstMeson}[2]{\ensuremath{\D_{\squark#1}^\ast(#2)^+}\xspace}
\newcommand{\DsstMesonNoMass}[1]{\ensuremath{\D_{\squark#1}^{\ast+}}\xspace}
\newcommand{\KstzMeson}[2]{\ensuremath{\kaon_{#1}^\ast(#2)^0}\xspace}
\newcommand{\DstzMeson}[2]{\ensuremath{\D_{#1}^\ast(#2)^0}\xspace}

\newcommand{\asymtight}[2]{%
  \ensuremath{%
    ^{+\vphantom{#2}\scriptscriptstyle #1}%
    _{-\vphantom{#1}\scriptscriptstyle #2}%
  }%
}

%% file: title-LHCb-PAPER.tex

\begin{titlepage}
\pagenumbering{roman}

\vspace*{-1.5cm}
\centerline{\large EUROPEAN ORGANIZATION FOR NUCLEAR RESEARCH (CERN)}
\vspace*{1.5cm}
\noindent
\begin{tabular*}{\linewidth}{lc@{\extracolsep{\fill}}r@{\extracolsep{0pt}}}
\ifthenelse{\boolean{pdflatex}}
{\vspace*{-1.5cm}\mbox{\!\!\!\includegraphics[width=.14\textwidth]{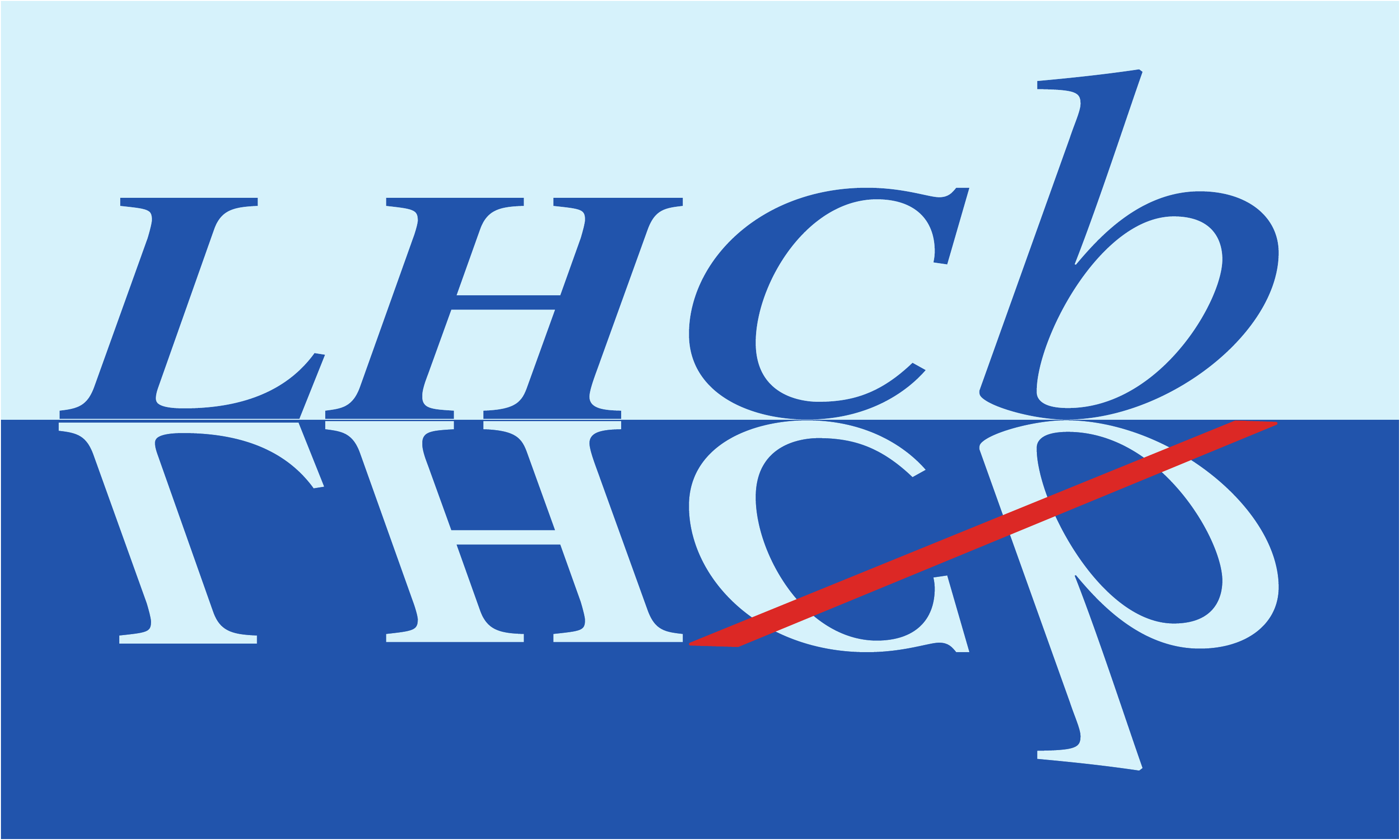}} & &}%
{\vspace*{-1.2cm}\mbox{\!\!\!\includegraphics[width=.12\textwidth]{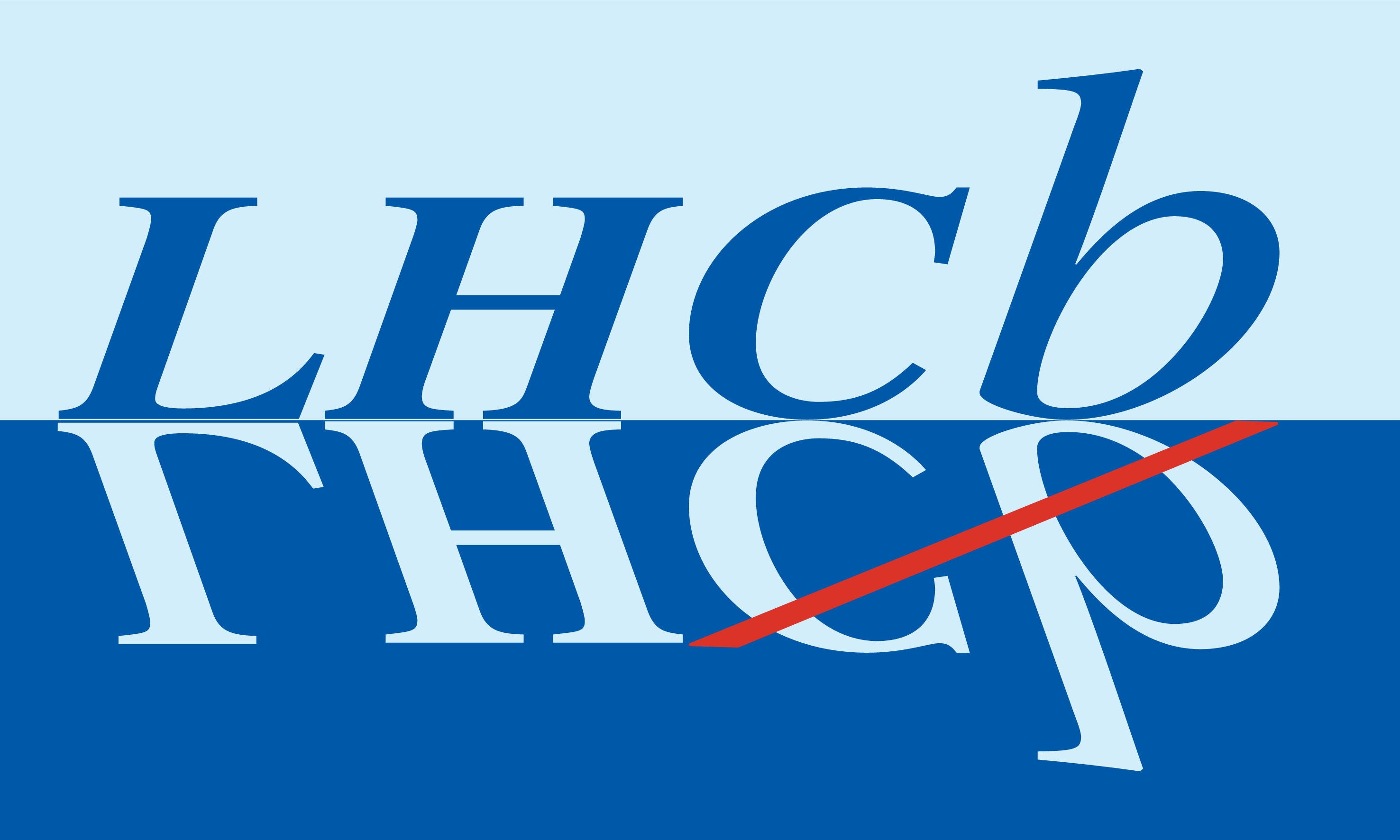}} & &}%
\\
 & & CERN-EP-2026-065 \\  
 & & LHCb-PAPER-2025-073 \\  
 & & April 23, 2026 \\ 
 & & \\
\end{tabular*}

\vspace*{4.0cm}

{\normalfont\bfseries\boldmath\huge
\begin{center}
  \papertitle 
\end{center}
}

\vspace*{2.0cm}

\begin{center}
\paperauthors\footnote{Authors are listed at the end of this paper.}
\end{center}

\vspace{\fill}

\begin{abstract}
  \noindent
A new  excited charm-strange meson is observed through an amplitude analysis of the full phase space of $B^0\to D^+ D^- K^+ \pi^-$ decays. The analysis is based on a proton-proton collision data sample collected by the \lhcb experiment at a center-of-mass energy $\sqs = 13\tev$, corresponding to an integrated luminosity of $5.4\invfb$.
The statistical significance of the new state exceeds $10$ standard deviations.
Its Breit--Wigner mass and width are measured to be
$m_0 = 2933^{+6}_{-5}(\text{stat})^{+4}_{-3}(\text{syst}) \,\text{MeV}
$
and  
$\Gamma_0 =
72^{+18}_{-12}(\text{stat})^{+\phantom{0}7}_{-10}(\text{syst}) \,\text{MeV}
$,
respectively, and its spin-parity quantum numbers are determined to be $J^P = 1^+$.
This new meson, denoted as $D_{s1}(2933)^+$, is a candidate for a $D_s(2P^{(\prime)}_{1})^+$ state.

\end{abstract}

\vspace*{2.0cm}

\begin{center}
  Submitted to
  Phys.~Rev.~Lett.
\end{center}

\vspace{\fill}

{\footnotesize 
\centerline{\copyright~\papercopyright. \href{\paperlicenceurl}{\paperlicence}.}}
\vspace*{2mm}

\end{titlepage}


\newpage
\setcounter{page}{2}
\mbox{~}
%
%
%
%

%% file: body.tex
The spectroscopy of charm-strange mesons provides a crucial environment for testing quantum chromodynamics~(QCD) in the nonperturbative regime~\cite{Guo2019CharmSpec,vanBeveren2021Review,Godfrey2015CharmSpec}.  
While the ground states and several low-lying \Ds excitations are established, a number of open questions remain.
A longstanding puzzle concerns the nature of the $\DsstMeson{0}{2317}$ and $\DsMeson{1}{2460}$ mesons~\cite{BaBar2003Ds2317,CLEO:2003ggt,BaBar:2003cdx,Belle2003Ds2460}, 
whose masses lie about 100\mev below quark-model expectations for $P$-wave excitations of the \Ds meson~\cite{Godfrey2015CharmSpec}.  
These features have motivated interpretations of these states  as $D^{(*)}K$ molecules and compact $c\bar{s}q\bar{q}$ tetraquark states~\cite{Guo2019CharmSpec,Mohler2013Ds2317LQCD,Bali2017DsLQCD,Ni2022DsSpectrum,Ni2023UnquenchedModel,Barnes:2003dj,Cheng:2003kg,Maiani:2004vq}.\footnote{$q$ denotes either $\uquark$ or $\dquark$ quark.} 
The recently observed $T^{*}_{c\bar{s}}(2900)^{0/++}$ tetraquark states~\cite{LHCb-PAPER-2022-026,LHCb-PAPER-2022-027} and $T_{c\bar{s}}(2327)^{0/++}$ tetraquark candidates~\cite{LHCb-PAPER-2024-033}, all with quark content $c\bar{s}q\bar{q}^{\prime}$ and belonging to isospin triplets, further support the  existence of singly charged $c\bar{s}q\bar{q}$ tetraquarks.
The discrepancy between the measured mass of the $\DsMeson{0}{2590}$ state and the quark-model predictions~\cite{LHCb-PAPER-2020-034,Godfrey2015CharmSpec} also suggests that further refinements to the quark-model description of the \Ds spectrum are required, for example through the inclusion of coupled-channel effects~\cite{Xie:2021Ds2590,Hao:2022CoupledChannel,Ortega:2022Ds2590,Yang:2022DsCoupled,Gao:2022Ds2590,Yang:2023CharmStrange} or mixing between the $1S$ and $2S$ states~\cite{Arifi:2022Mixing}.
Furthermore, the majority of higher excitations such as $2P$, $1D$ and $2S$ remain unobserved or poorly characterized ~\cite{Moir2013LQCD,Ni2022DsSpectrum,Ni2023UnquenchedModel}. 
Clarifying the nature of the unexpectedly low-lying charm-strange states and completing the map of high-mass $\Ds$ excitations is therefore central to understanding how QCD dynamics shape the charm-strange meson spectrum~\cite{Guo2019CharmSpec,vanBeveren2021Review}.

Multibody $B$-meson decays provide an excellent environment to study $\Ds$ resonances in the $D^{(*)} K^{(*)}$ system~\cite{BaBar2006DsJinB,Belle2003Ds2460}. 
These processes tend to have a large final-state phase space, giving access to a wide range of intermediate states.  Moreover, where their decay kinematics are fully reconstructed, the use of amplitude-analysis techniques enables precise determinations of the properties of the resonant structures.
The \BdToDDKpi decay is a representative example of this class of decays.\footnote{The inclusion of charge-conjugate processes is implied throughout this Letter, and natural units with $c = \hbar = 1$ are used.} In this mode, excited \Ds\ mesons with any spin-parity ($J^P$) combination except for $J^P=0^{+}$ can, if kinematically allowed, be reconstructed in the $\Dp\Kp\pim$ three-body system.  This process thereby offers opportunities to observe significantly more states than in the more extensively studied $D K$ spectra, which only give access to natural spin-parity states~($0^{+}$, $1^{-}$, $2^{+}$, ...).

The \BdToDDKpi decay has been previously studied by the \lhcb collaboration in a restricted phase-space region with $m(\Kp\pim)<750\mev$, and resulted in the observation of the $\DsMeson{0}{2590}$ state~\cite{LHCb-PAPER-2020-034}.
In this Letter, an amplitude analysis of the full phase space is presented using the proton-proton~(\proton\proton) collision data sample collected by the \lhcb experiment at center-of-mass energy \mbox{$\sqs = 13\tev$} in  2016--2018, corresponding to an integrated luminosity of \mbox{$5.4\invfb$}, which is the same as that in the aforementioned analysis. 

The LHCb detector~\cite{LHCb-DP-2008-001,LHCb-DP-2014-002} is a single-arm forward spectrometer covering the pseudorapidity range $2 < \eta < 5$, designed for studies of particles containing $b$ or $c$ quarks. The online event selection is performed by a trigger~\cite{LHCb-DP-2012-004}, which consists of a hardware stage, based on information from the calorimeter and muon systems, followed by a software stage, which applies a full event reconstruction.
Triggered data further undergo a centralized, offline processing step to deliver physics-analysis-ready data across the entire \lhcb physics program~\cite{Stripping}.
 The momentum scale is calibrated using samples of \decay{\jpsi}{\mup\mun} and \decay{\Bp}{\jpsi\Kp} decays, collected concurrently with the data sample employed in this analysis~\cite{LHCb-PAPER-2012-048,LHCb-PAPER-2013-011}. Simulated samples, produced with the software packages described in Refs.~\cite{Sjostrand:2007gs,Lange:2001uf,Golonka:2005pn,Allison:2006ve,*Agostinelli:2002hh,LHCb-PROC-2011-006}, are used to model the effects of the detector acceptance and the applied selection requirements. In these samples, the \BdToDDKpi decays are generated uniformly in the phase space. The simulated samples are corrected to account for known data-simulation differences in the \Bz production kinematics and detector occupancy, as well as track-reconstruction, particle-identification (PID) and trigger efficiencies.

The \BdToDDKpi decay is reconstructed using the $\Dpm\to\Kmp\pipm\pipm$ decay mode.
All final-state kaons and pions are required to have PID information consistent with their respective mass hypothesis~\cite{LHCb-DP-2018-001}, and to be inconsistent with originating from any primary $\proton\proton$ collision vertex~(PV). 
To suppress backgrounds arising from the repeated use of track segments, the opening angle between any pair of tracks is required 
to be larger than $0.5\mrad$.
The \Dpm candidates must have good vertex-fit quality and a reconstructed mass within $\pm25\mev$ of the known value~\cite{PDG2024}. 
The \Bz candidates are required to have both good vertex-fit quality and a reconstructed momentum vector aligned with the vector from the associated PV to the decay vertex. The associated PV is defined as the PV that fits best to the flight direction of the \Bz candidates.
Background from \Bz decays to the same final state as the signal but lacking at least one
charm intermediate state is further suppressed by requiring a displacement between the \Bd and \Dpm decay vertices.
A kinematic fit is applied to the full decay chain to improve the \Bd mass resolution, constraining the \Bd candidate to originate from the associated PV and the \Dpm masses to their known values~\cite{PDG2024}. A second kinematic fit, in which the \Bz mass is additionally constrained to its known value, is used to compute the variables entering the amplitude analysis.

To suppress background from random combinations of final-state tracks, referred to hereafter as combinatorial background, a multivariate classifier based on the gradient boosted decision tree~(GBDT) classifier~\cite{GBDT,Breiman,AdaBoost}, as implemented in the TMVA toolkit~\cite{Hocker:2007ht,TMVA4}, is employed. 
The input variables are the difference in the vertex-fit \chisq of a given PV reconstructed with and without the track(s) under consideration~(\chisqip), the PID information of all final-state particles, and the vertex-fit quality, the transverse momentum~(\pt) and the \chisqip of the \Bz candidate. 
The classifier is trained using simulated \BdToDDKpi decays as the signal sample and data candidates in the range $5380 < m(\Dp\Dm\Kp\pim) < 5530\mev$ as the background sample.
The optimal working point on the GBDT response is chosen by maximizing the expected signal significance, \mbox{$S/\sqrt{S+B}$}, where $S$ and $B$ denote the signal and background yields, respectively, in the \Bz signal region defined as the \(\pm 20\mev\) window around the known \Bz mass~\cite{PDG2024}. Some candidates passing all selection criteria contain multiple candidates. In such cases the candidate with the best kinematic fit \chisq is retained. 

\begin{figure}
    \centering
    \includegraphics[width=0.49\linewidth]{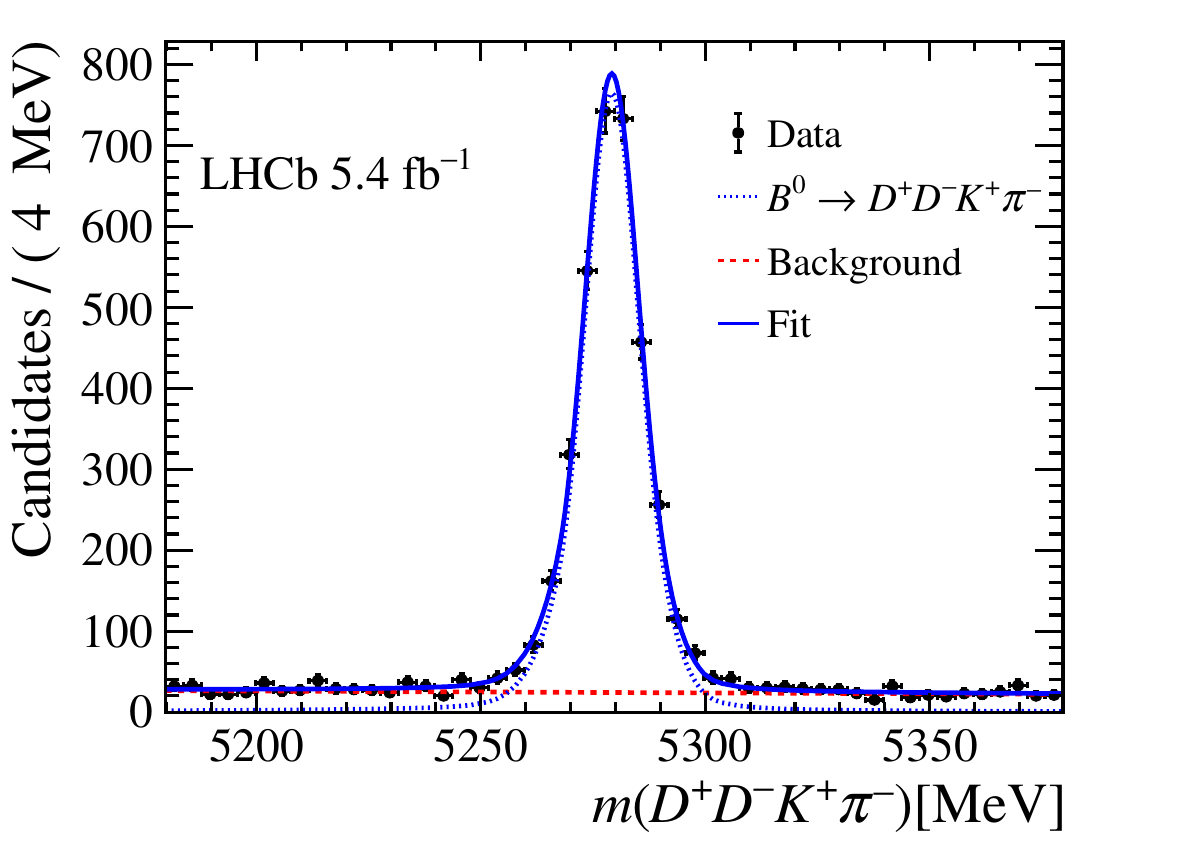}
    \includegraphics[width=0.49\linewidth]{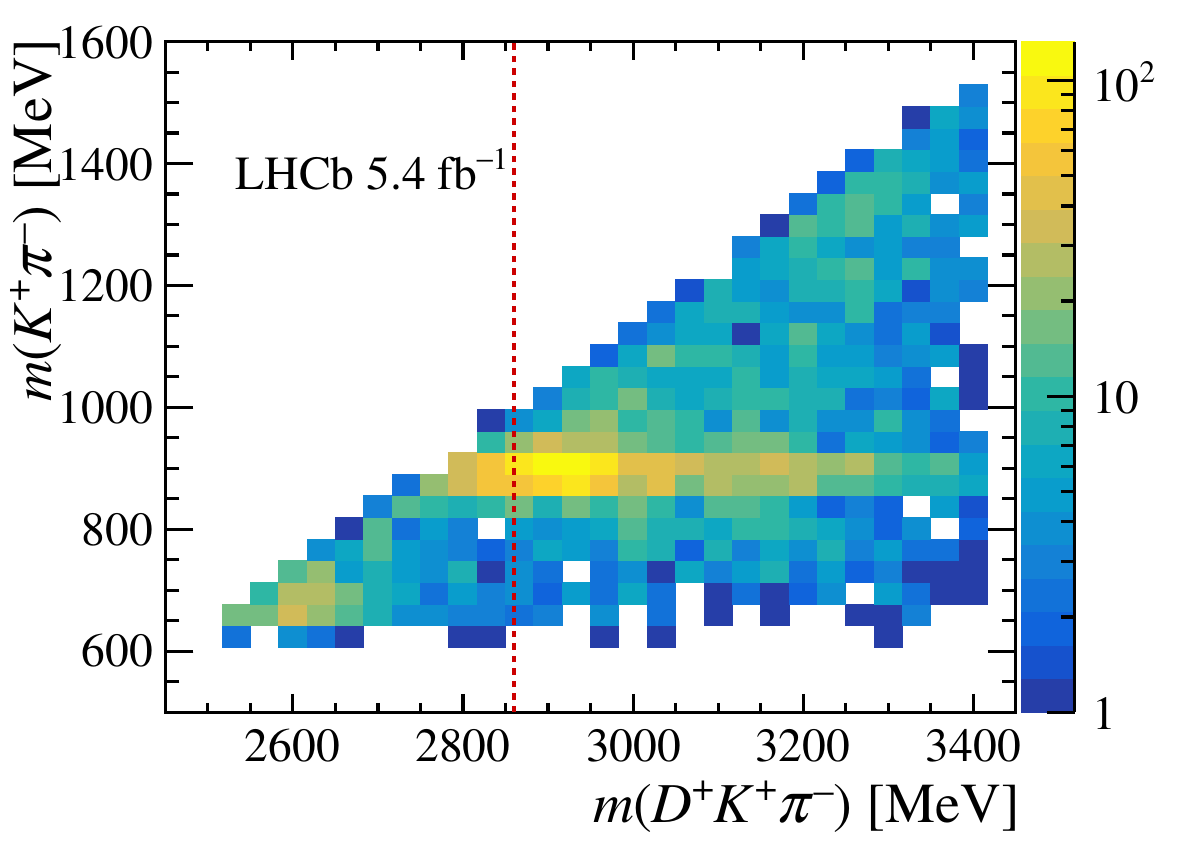}
    \caption{(Left) Mass distribution of the selected \Bz candidates with the fit result also shown. 
    (Right) Two-dimensional background-subtracted mass distribution of $m(\Kp\pim)$ versus $m(\Dp\Kp\pim)$. 
    The red dashed line indicates the peak position of the \DsstMeson{1,3}{2860} states.
    }
    \label{fig:massfit_and_Dalitz}
\end{figure}

An extended unbinned maximum-likelihood fit is performed to the mass distribution of the \Bz candidates, as shown in Fig.~\ref{fig:massfit_and_Dalitz}. 
The fit model is composed of a signal component, described by the sum of two Crystal Ball functions~\cite{Skwarnicki:1986xj} sharing a common peak position and power-law tails on both sides of the peak, with tail parameters fixed from simulation, and a combinatorial background component modeled with an exponential function.
Backgrounds arising from the misidentification of a final-state pion, kaon, or proton in \bquark-hadron decays or from partially reconstructed decays with missing final-state particles have been investigated in the previous analysis~\cite{LHCb-PAPER-2020-034} and found to be negligible.
The fitted signal yield is \mbox{$\val{massfit_yield}\pm\val{massfit_uncert}$}, where the uncertainty is statistical only. 
The resulting two-dimensional $\Kp\pim$ versus $\Dp\Kp\pim$ invariant-mass distribution, with the background subtracted statistically using the \sPlot method~\cite{Pivk:2004ty}, is also shown in Fig.~\ref{fig:massfit_and_Dalitz}. 
A cluster of signal decays near the phase-space edge around $\mDKpi \approx 2600\mev$ corresponds to the $\DsMeson{0}{2590}$ state previously observed in this channel~\cite{LHCb-PAPER-2020-034}. Moreover, a distinct enhancement in the $\Kstar(892)^0$ region at ${\mDKpi \approx 2950\mev}$, exceeding the mass of the previously observed \DsstMeson{1,3}{2860} state~\cite{LHCb-PAPER-2014-035}, suggests the potential existence of a new excited \Ds contribution. To investigate this structure, an amplitude analysis over the full five-dimensional phase space is performed.  

The total \BdToDDKpi decay amplitude $\mathcal{M}$ is constructed as a coherent sum of 
partial amplitudes corresponding to intermediate resonant or nonresonant components within the isobar-model framework~\cite{Lindenbaum:1957ec}, where the four-body decay is disentangled into a sequence of two-body decays whose amplitudes are formulated in the helicity formalism~\cite{Jacob:1959at,Chung:1993da}.  Two general decay topologies are considered:
\begin{description}
    \item[Cascade:] $\Bz\to P_1 T_1,\, T_1\to P_2 T_2,\, T_2\to P_3 P_4$, 
    \item[Quasi-two-body:] $\Bz\to R_1 R_2,\, R_1\to P_1 P_2,\, R_2\to P_3 P_4$, 
\end{description}
where $P_i \in \{D^+, D^-, K^+, \pi^-\}$, and $T_i$ ($R_i$) denote intermediate resonant or nonresonant components
appearing in the cascade (quasi-two-body) topology.
The amplitudes associated with the two topologies are
\begin{equation}
\begin{aligned}
    \mathcal{M}^{T_{1}, T_{2}}_{\text{casc}}  &=  
    \sum_{\lambda_{T_{2}}^{T_{1}}} 
   \mathcal{H}^{{T_{1}},{T_2}}_{\lambda_{T_{2}}^{T_{1}} }
   \,
   e^{i \lambda_{T_{2}}^{T_{1}} \phi^{P_3}_{T_2}} 
   \times
   d^{J_{T_{1}}}_{0,\lambda_{T_{2}}^{T_{1}}}(\theta_{T_{1}}^{T_2})
   \, 
   d^{J_{T_{2}}}_{\lambda_{T_{2}}^{T_{1}},0}(\theta_{T_{2}}^{P_3})
   \times
   \mathcal{A}_{T_1}(m_{{P_{2}}{P_3}{P_4}})
   \mathcal{A}_{T_2}(m_{{P_3}{P_4}}), \\
    \mathcal{M}^{R_{1},R_{2}}_{\text{quas }} &= 
    \sum_{\lambda_{R_{1}}^{\Bz}}
     \mathcal{H}^{R_{1}, R_{2}}_{\lambda_{R_{1}}^{\Bz}} 
     \,
     e^{i\lambda_{R_{1}}^{\Bz} (\phi^{P_1}_{R_1}+\phi^{P_3}_{R_2})}  
     \times
     d^{J_{R_{1}}}_{\lambda_{R_{1}}^{\Bz},0}(\theta_{R_1}^{P_1}) 
     \,
     d^{J_{R_{2}}}_{\lambda_{R_{1}}^{\Bz},0}(\theta_{R_2}^{P_3}) 
     \times
     \mathcal{A}_{R_1} (m_{P_1P_2}) 
     \mathcal{A}_{R_2} (m_{P_3P_4}),
\label{eq:amp}
\end{aligned}
\end{equation}
where the angular dependence is described by two Wigner $d$-function terms,
evaluated at the polar angles $\theta_Y^X$, multiplied by an exponential term that depends
on the azimuthal angles $\phi_Y^X$.  The angles are defined from the direction of the
momentum of the decay product $X$ in spherical coordinates in the rest frame of the
decaying particle $Y$, following the convention of Ref.~\cite{LHCb-PAPER-2015-029}.  The
spin-parity of $Y$ is denoted as $J_Y$, and the helicity of $X$ in the $Y$ rest frame as
$\lambda_{X}^{Y}$.  The complex helicity couplings
$\mathcal{H}^{{T_{1}},{T_2}}_{\lambda_{T_{2}}^{T_{1}} }$ and
$\mathcal{H}^{R_{1}, R_{2}}_{\lambda_{R_{1}}^{\Bz}}$ characterize the magnitude and phase of
the contributing amplitudes. 
The explicit form of the mass-dependent terms $\mathcal{A}_Y(m)$, which is a function of
the invariant mass~($m$) of the involved final-state particles, varies according to
the component under consideration.  
Generally, 
resonances are described by a relativistic Breit--Wigner~(RBW) function,
with the meson radius in the Blatt--Weisskopf factor set to $r=3\gev^{-1}$~\cite{PDG2024}.
The mass-dependent width $\Gamma(m)$ in the RBW function is calculated from the  sum of partial widths of all kinematically allowed decay channels~($c$) of the resonance, expressed as
\begin{equation}
\Gamma(m)
=
\sum_c \Gamma_{c0}
\frac{I_c(m)}{I_c(m_0)},
\qquad
I_c(m)
=
\int_{m} 
|\mathcal{M}_c(\Phi_c)|^2 \, \deriv\Phi_c \, .
   \label{eq:width}
\end{equation}
Here, $\Phi_c$ denotes the phase-space variables associated with the decay channel $c$ of the resonance, while the matrix element $\mathcal{M}_c$ is parametrized analogously to the $\Bz$ decay amplitude in Eq.~\eqref{eq:amp}.
The integral is performed over the accessible phase space, whose boundary depends on the invariant mass of the final-state particles involved, as indicated by the subscript of the integral symbol. The Breit-Wigner mass is denoted by $m_0$, and the Breit-Wigner width is given by $\Gamma_0 = \Gamma(m_0) = \sum_c \Gamma_{c0}$, where $\Gamma_{c0}$ is the partial width of channel $c$.
For a two-body decay, the width $\Gamma(m)$ can be evaluated analytically as shown in Ref.~\cite{PDG2024}, while for multibody decays the integral is computed numerically. 

With the decay amplitude defined, the signal probability density function~(PDF) is written as
\begin{equation}
\mathcal{P}_{\mathrm{sig}}(m,\Omega\mid\vec{\omega})
= \frac{1}{I(\vec{\omega})}
\big|\mathcal{M}(m,\Omega\mid\vec{\omega})\big|^2 
\Phi(m)\,\epsilon(m,\Omega),
\label{eq:sig_pdf}
\end{equation}
where $(m,\Omega)$ denotes the invariant-mass and angular variables that specify a point in phase space, $\Phi(m)$ is the four-body phase-space density, and $\vec{\omega}$ is the set of fit parameters that generally include the amplitude couplings and mass-lineshape parameters. 
The normalization factor $I(\vec{\omega})$ is obtained by summing $|\mathcal{M}|^2$ values calculated over simulated \BdToDDKpi decays generated uniformly in phase space and passing all selections. The term $\epsilon(m,\Omega)$ accounts for the efficiency variation across phase space, and is determined from the same simulation sample used to evaluate $I(\vec{\omega})$ with the kernel density estimation~(KDE) technique~\cite{Silverman1986}, which provides a smooth, nonparametric description of multidimensional distributions. To reduce the effective dimensionality of the KDE, the full five-dimensional four-body phase space is factorized into invariant-mass and angular subspaces, as the correlations between them are verified to be negligible.

Subsequently, the total PDF is constructed as
\begin{equation}
P_{\rm tot}(m, \Omega \mid \vec{\omega}) = (1-\beta)\, \mathcal{P}_{\mathrm{sig}}(m, \Omega\mid \vec{\omega}) + \beta\, \mathcal{P}_{\mathrm{bkg}}(m, \Omega),
\label{eq:pdf}
\end{equation}
where $\beta=0.068\pm0.002$ represents the background fraction in the $\Bz$ signal region obtained from the \Bz mass fit, and $\mathcal{P}_{\mathrm{bkg}}(m,\Omega)$ denotes the background PDF obtained from candidates in the \Bz mass sideband using the KDE technique in the same way as for the efficiency determination. The total PDF is used to construct the unbinned negative log-likelihood~(NLL) function, which is minimized during the amplitude fit to find the best model and to determine the values of the parameters of interest.

The development of the amplitude model begins by identifying all established conventional resonances~\cite{PDG2024} that may contribute to the \BdToDDKpi decay, classified into the two categories introduced above:

\begin{description}

\item[Cascade:] the excited charm-strange mesons $\DsMeson{1}{2536}$, $\DsMeson{0}{2590}$, $\DsstMeson{1}{2700}$, $\DsstMeson{1}{2860}$ and $\DsstMeson{3}{2860}$ decaying to $\Dp\Kp\pim$ via either the $\Kstarz$ resonances $\KstzMeson{}{892}$, $\KstzMeson{1}{1410}$, $\KstzMeson{2}{1430}$ and $(\Kp\pim)_S$; or the $\Dstarz$ resonances $\DstzMeson{2}{2460}$, $\DstzMeson{1}{2600}$, and $(\Dp\pim)_S$. Here, $(\Kp\pim)_S$ denotes the $S$-wave component including \KstzMeson{0}{1430}  and nonresonant contributions, while $(\Dp\pim)_S$ represents a purely nonresonant 
$S$-wave contribution. Nonresonant three-body decays $\Bz\to \Dstarz \Dm \Kp$ and $\Bz\to \Dp\Dm \Kstarz$ are also considered, with the $\Dstarz\Kp$ and $\Dp \Kstarz$ systems, collectively denoted as $\Ds\,\text{NR}$, allowed to have $J^P=0^-$ and $1^+$.

\item[Quasi-two-body:] the charmonium and charmonium-like states $\psi(3770)$, $\chiczero(3915)$, $\chictwo(3930)$, $\psi(4040)$, $\psi(4160)$, $\psi(4230)$, $\psi(4360)$ and $\psi(4415)$ decaying into $\Dp\Dm$, accompanied by either $(\Kp\pim)_S$ or $\Kstarz$ resonant contributions.
\end{description}
These combinations are included only when they are kinematically allowed and permitted by angular momentum conservation and, for strong decays, parity conservation. 
Mass lineshapes for the \Kstarz, \Dstarz and charmonium states are modeled using the RBW function, with mass-dependent widths $\Gamma(m)$ calculated using only $\Kp\pim,\Dp\pim$ and $\Dp\Dm$ decays, respectively.
The $\DsMeson{1}{2536}$, $\DsMeson{0}{2590}$ and $\DsstMeson{3}{2860}$ lineshapes are described by RBW functions with multichannel widths that include all dominant decay modes, with the branching fractions of the $\DsMeson{1}{2536}$ and $\DsMeson{0}{2590}$ states set to known values~\cite{PDG2024,LHCb-PAPER-2020-034} and those of the $\DsstMeson{3}{2860}$ state to theoretical predictions~\cite{Godfrey2015CharmSpec}.
For the $\DsstMeson{1}{2700}$ and $\DsstMeson{1}{2860}$ states, their substantial overlap and identical  quantum numbers of $J^P=1^-$ make a coherent sum of RBW amplitudes inadequate to capture their interference, and thus a quasi-model-independent~(MI) spline parametrization is used for the lineshape, with the amplitude free to vary at knot points $[2.65,\,2.70,\,2.80,\,2.85,\,2.90,\,2.95,\,3.00,\,3.10,\,3.20,\,3.30,\,3.40]\gev$. 
The $(\Kp\pim)_S$ amplitude is described by the LASS parametrization~\cite{lass}.
The $(\Dp\pim)_S$ and $\Ds\, \text{NR}$ components are modeled by an exponential function $\mathcal{A}(m) = \exp{\left(-(\alpha+i\lambda)m^2\right)}$, with parameters $\alpha$ and $\lambda$ free in the fit.

An initial amplitude fit is attempted using only the established conventional amplitudes described above. The $\Bz\to\DsMeson{0}{2590}\Dm$, $\DsMeson{0}{2590}\to\Dp(\Kp\pim)_{S}$ amplitude is chosen as the reference, with its helicity coupling fixed to unity. The masses and widths of all resonances are fixed to their known values~\cite{PDG2024}, except for those of the $\DsMeson{0}{2590}$ state~\cite{LHCb-PAPER-2020-034} which are allowed to float. 
Additional free parameters include the helicity couplings of the remaining amplitudes, the knot-point amplitudes associated with the spline lineshape, and the parameters of the nonresonant components.

\begin{figure}[tp]
    \centering
    \includegraphics[width=0.49\linewidth]{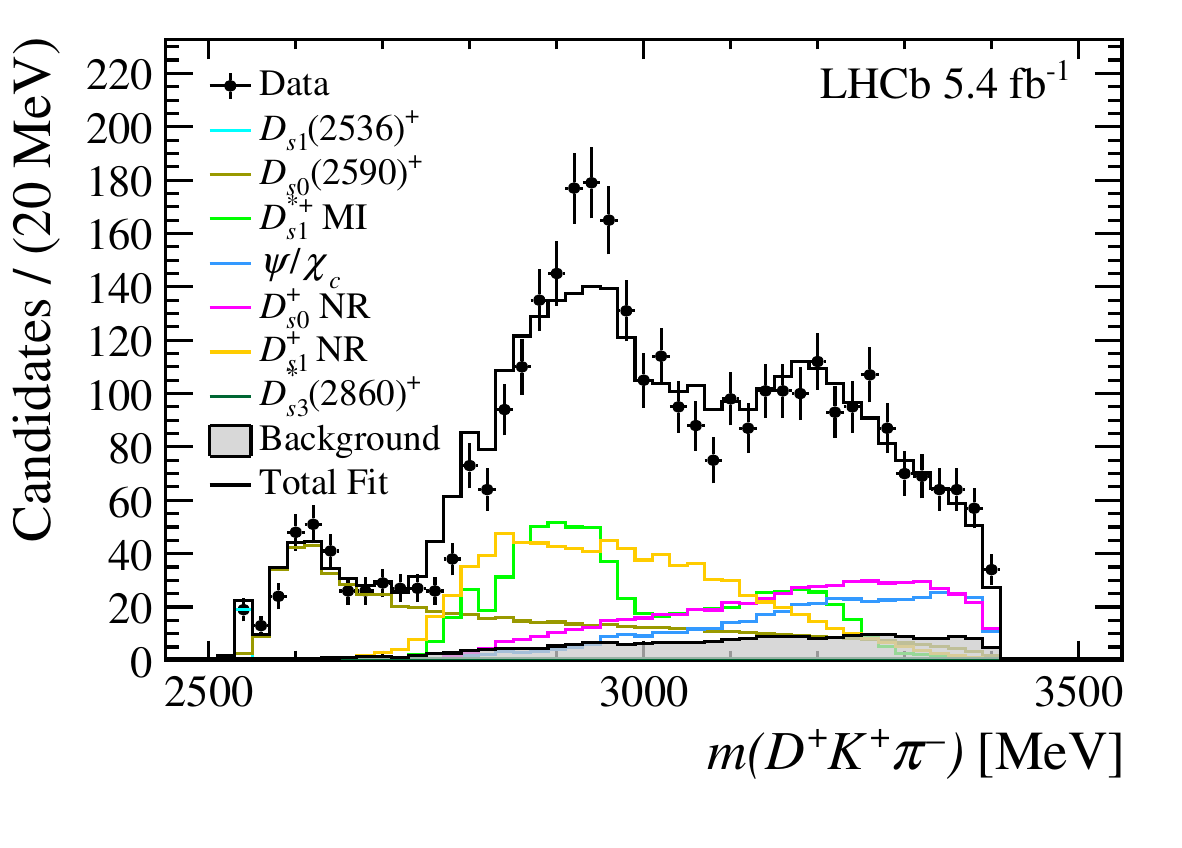}
    \includegraphics[width=0.49\linewidth]{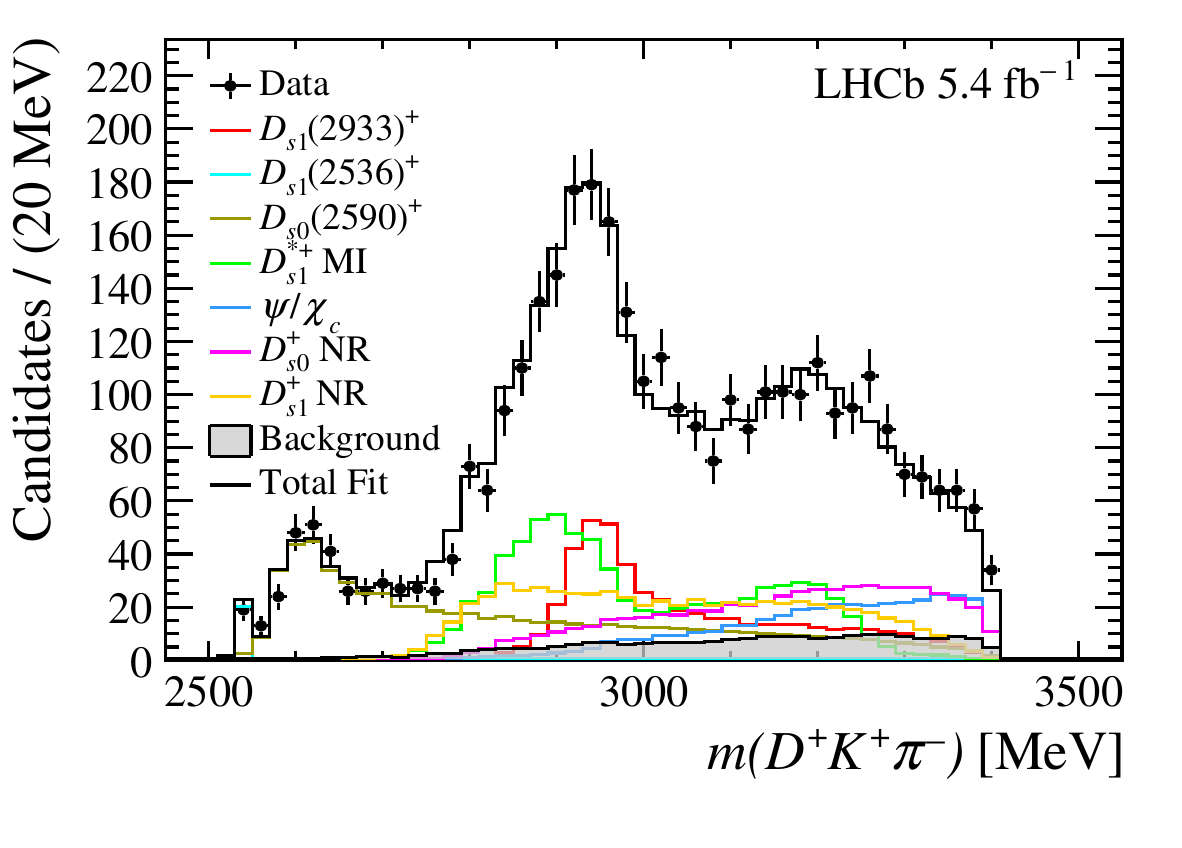}
    \caption{
    Distributions of $\mDKpi$ for \BdToDDKpi candidates in the signal region with fit results from the (left) initial and (right) baseline models. The components labeled MI and NR correspond to the quasi-model-independent amplitude for $J^P=1^-$ \Dsp states and nonresonant amplitudes, respectively. The amplitudes of the included charmonium states are summed and denoted collectively as $\psi/\chi_c$.
    }
    \label{fig:Model_mDKpi}
\end{figure}

This initial model does not describe the data well, with a significant discrepancy observed at $\mDKpi \approx 2950\mev$, as shown in Fig.~\ref{fig:Model_mDKpi}~(left). An additional \Ds resonance is therefore introduced in the model. It is modeled by an RBW function with its mass and width left free in the fit. Its dominant $D^{(*)}K^{(*)}$ channels are included, with their branching fractions fixed to the values predicted in Ref.~\cite{Godfrey2015CharmSpec}.
Several $J^P$ hypotheses, including $0^-$, $1^+$ and $2^\pm$, are tested, among which the $J^P=1^+$ assignment yields the lowest NLL value and is thus favored.
The statistical significance of the new \Ds resonance is determined to be over 10 standard deviations from a comparison of the NLL values obtained from fits with and without this component using Wilks' theorem~\cite{Wilks:1938dza}. 
Alternative $J^P$ hypotheses are tested against the $J^P=1^+$ assignment and are rejected with significances exceeding five standard deviations. No other statistically significant \Ds states are found.

The baseline amplitude model is obtained by subsequently removing amplitudes with statistical significance less than 3 standard deviations from the initial model with the new \Ds state.
The retained components are listed in Table~\ref{tab:baseline_ff}, and the fit projection in the $\mDKpi$ distribution is shown in Fig.~\ref{fig:Model_mDKpi}~(right).
The significance of the new \Ds state is largely unaffected by this procedure. 
Its mass and width are measured to be
\begin{equation*}
\begin{aligned}
m_0 &=
\val{newDs_Mass_mev}\,
^{+\val{newDs_Mass_stat_upper_mev}}_{-\val{newDs_Mass_stat_lower_mev}}\stat\,
^{+\val{newDs_Mass_syst_upper_mev}}_{-\val{newDs_Mass_syst_lower_mev}}\syst
\mev, \\
\Gamma_0 &=
\val{newDs_Width_mev}\,
^{+\val{newDs_Width_stat_upper_mev}}_{-\val{newDs_Width_stat_lower_mev}}\stat\,
^{+\phantom{0}\val{newDs_Width_syst_upper_mev}}_{-\val{newDs_Width_syst_lower_mev}}\syst
\mev,
\end{aligned}
\end{equation*}
respectively, where the first uncertainty is statistical and the second systematic.
Since its spin-parity is $J^P=1^+$, the new state is denoted as \DsMeson{1}{\val{newDs_Mass_mev}}. 
The pole mass and width of the $\DsMeson{0}{2590}$   are found to be
\mbox{$
\val{Ds2590_pole_Mass_mev}\,
^{+\val{Ds2590_pole_Mass_stat_upper_mev}}_{-\val{Ds2590_pole_Mass_stat_lower_mev}}\stat\,
^{+\val{Ds2590_pole_Mass_syst_upper_mev}}_{-\val{Ds2590_pole_Mass_syst_lower_mev}}\syst
\mev
$}
and  
\mbox{$
\val{Ds2590_pole_Width_mev}\,
^{+\phantom{0}\val{Ds2590_pole_Width_stat_upper_mev}}_{-\val{Ds2590_pole_Width_stat_lower_mev}}\stat\,
^{+\phantom{0}\val{Ds2590_pole_Width_syst_upper_mev}}_{-\val{Ds2590_pole_Width_syst_lower_mev}}\syst
\mev
$}, respectively, consistent with the values reported in the previous analysis~\cite{LHCb-PAPER-2020-034}.
Fit fractions, defined as in Ref.~\cite{LHCb-PAPER-2015-029}, are reported in Table~\ref{tab:baseline_ff} for all components in the amplitude model.

\begin{table}[tp]
\centering
\caption{Fit fractions of the components in the baseline amplitude model.
The first uncertainty is statistical and the second systematic. The symbol $\DsstMesonNoMass{1}\,\text{MI}$ denotes the quasi-model-independent spline shape accounting for the \DsstMeson{1}{2700} and \DsstMeson{1}{2860} states, while $\DsMesonNoMass{0}\,\text{NR}$ and $\DsMesonNoMass{1}\,\text{NR}$ are \Dsp nonresonant amplitudes with spin-parity $0^-$ and $1^+$ , respectively. }
\renewcommand{\arraystretch}{1.1}
\begin{tabular}{lc}
\hline
Component & Fit fraction (\%) \\
\hline
$\Bz\to\Dm\, \DsMeson{1}{2536},\, \DsMeson{1}{2536}\to\Kp\, (\Dp\pim)_S$ & 1.0\,\asymtight{0.1}{0.3}\,\asymtight{0.6}{0.4} \\
$\Bz\to\Dm\, \DsMeson{0}{2590},\, \DsMeson{0}{2590}\to\Dp\, (\Kp\pim)_S$ & 21.5\,\asymtight{0.4}{2.1}\,\asymtight{2.5}{0.6} \\
$\Bz\to\Dm\, \DsMeson{1}{2933},\, \DsMeson{1}{2933}\to\Kp\, (\Dp\pim)_S$ & 2.4\,\asymtight{0.5}{0.7}\,\asymtight{0.4}{1.3} \\
$\Bz\to\Dm\, \DsMeson{1}{2933},\, \DsMeson{1}{2933}\to\Dp\, \KstzMeson{1}{892}$ & 8.9\,\asymtight{1.8}{2.1}\,\asymtight{2.4}{1.8} \\
$\Bz\to\Dm\, \DsMeson{1}{2933},\, \DsMeson{1}{2933}\to\Kp\, \DstzMeson{2}{2460}$ & 7.0\,\asymtight{6.0}{3.0}\,\asymtight{3.0}{3.7} \\
$\Bz\to\Dm\, \DsstMesonNoMass{1}\,\text{MI},\, \DsstMesonNoMass{1}\,\text{MI} \to\Dp\, \KstzMeson{1}{892}$ & 18.0\,\asymtight{0.9}{2.4}\,\asymtight{0.4}{2.0} \\
$\Bz\to\Dm\, \DsstMesonNoMass{1}\,\text{MI},\, \DsstMesonNoMass{1}\,\text{MI} \to\Kp\, \DstzMeson{2}{2460}$ & 4.7\,\asymtight{0.8}{0.5}\,\asymtight{0.4}{0.3} \\
$\Bz\to\Dm\, \DsMesonNoMass{0}\,\text{NR},\, \DsMesonNoMass{0}\,\text{NR} \to\Dp\, \KstzMeson{1}{892}$ & 8.4\,\asymtight{2.2}{0.8}\,\asymtight{1.9}{0.9} \\
$\Bz\to\Dm\, \DsMesonNoMass{0}\,\text{NR},\, \DsMesonNoMass{0}\,\text{NR} \to\Dp\, \KstzMeson{1}{1410}$ & 1.6\,\asymtight{1.1}{0.3}\,\asymtight{0.7}{0.9} \\
$\Bz\to\Dm\, \DsMesonNoMass{0}\,\text{NR},\, \DsMesonNoMass{0}\,\text{NR} \to\Dp\, \KstzMeson{2}{1430}$ & 1.0\,\asymtight{0.5}{0.3}\,\asymtight{0.9}{0.1} \\
$\Bz\to\Dm\, \DsMesonNoMass{0}\,\text{NR},\, \DsMesonNoMass{0}\,\text{NR} \to\Kp\, \DstzMeson{2}{2460}$ & 1.9\,\asymtight{0.5}{0.6}\,\asymtight{0.3}{0.9} \\
$\Bz\to\Dm\, \DsMesonNoMass{0}\,\text{NR},\, \DsMesonNoMass{0}\,\text{NR} \to\Kp\, \DstzMeson{1}{2600}$ & 1.5\,\asymtight{0.8}{0.3}\,\asymtight{0.4}{0.4} \\
$\Bz\to\Dm\, \DsMesonNoMass{1}\,\text{NR},\, \DsMesonNoMass{1}\,\text{NR} \to\Dp\, \KstzMeson{1}{892}$ & 10.0\,\asymtight{2.2}{1.1}\,\asymtight{1.8}{1.2} \\
$\Bz\to\Dm\, \DsMesonNoMass{1}\,\text{NR},\, \DsMesonNoMass{1}\,\text{NR} \to\Kp\, \DstzMeson{2}{2460}$ & 6.0\,\asymtight{6.5}{3.0}\,\asymtight{2.6}{4.4} \\
$\Bz\to\Dm\, \DsMesonNoMass{1}\,\text{NR},\, \DsMesonNoMass{1}\,\text{NR} \to\Kp\, \DstzMeson{1}{2600}$ & 2.5\,\asymtight{0.6}{0.5}\,\asymtight{0.6}{0.4} \\
$\Bz\to(\Kp\pim)_S\, \chiczero(3915),\chiczero(3915)\to\Dp\,\Dm$ & 6.7\,\asymtight{0.1}{2.2}\,\asymtight{0.4}{1.1} \\
$\Bz\to\KstzMeson{1}{892}\, \chiczero(3915),\chiczero(3915)\to\Dp\,\Dm$ & 0.9\,\asymtight{0.5}{0.3}\,\asymtight{0.3}{0.1} \\
$\Bz\to\KstzMeson{1}{892}\, \chictwo(3930),\chictwo(3930)\to\Dp\,\Dm$ & 0.7\,\asymtight{0.4}{0.3}\,\asymtight{0.2}{0.2} \\
$\Bz\to\KstzMeson{1}{892}\, \psi(3770),\, \psi(3770)\to\Dp\,\Dm$ & 2.3\,\asymtight{0.5}{0.4}\,\asymtight{0.4}{0.5} \\
$\Bz\to\KstzMeson{2}{1430}\, \psi(3770),\, \psi(3770)\to\Dp\,\Dm$ & 0.9\,\asymtight{0.4}{0.3}\,\asymtight{0.0}{0.3} \\
$\Bz\to\KstzMeson{1}{892}\, \psi(4040),\, \psi(4040)\to\Dp\,\Dm$ & 0.9\,\asymtight{0.6}{0.3}\,\asymtight{0.7}{0.3} \\
$\Bz\to\KstzMeson{1}{892}\, \psi(4160),\, \psi(4160)\to\Dp\,\Dm$ & 0.9\,\asymtight{0.7}{0.1}\,\asymtight{1.8}{0.3} \\
\hline
Total & $109.7\asymtight{12.4}{5.4}\,\asymtight{7.5}{9.5}$ \\\hline
\end{tabular}
\label{tab:baseline_ff}
\end{table}

Systematic uncertainties on the measured parameters are summarized in Tables~\ref{tab:paras_summary} and \ref{tab:ff_summary} of the End Matter.
The dominant uncertainties arise from modeling of the resonance lineshapes, and are assessed by replacing the baseline parametrizations with other commonly used descriptions of the same states. For instance, the $(\Kp\pim)_S$ component is described using the LASS parametrization~\cite{lass} in the baseline model, and is replaced by a $K$-matrix formalism with parameters taken from the FOCUS experiment~\cite{FOCUS:2007mcb} when evaluating the associated systematic uncertainty.
Other sources include imperfections in the corrections applied to simulated samples; 
limited size of input samples and the choice of hyperparameters in the KDE used for efficiency and background parameterizations; 
uncertainties from fixed resonance parameters and the Blatt--Weisskopf barrier radius;
inclusion of additional potential contributions from known resonances, namely $\DsstMeson{3}{2860}$, $T_{\cquark\cquarkbar1}(3900)^+$, $T_{\cquark\cquarkbar1}(4430)^+$, $T_{\cquarkbar\squarkbar0}^*(2870)^0$, $T_{\cquarkbar\squarkbar1}^*(2900)^0$, $T_{\cquark\squarkbar0}^*(2900)^0$, $\psi(4230)$, $\psi(4360)$, and $\psi(4415)$; 
the background fraction determined from the \Bz mass fit;
the treatment of backgrounds with only one or zero charm resonances and events with multiple \Bz candidates;
and potential intrinsic biases in the amplitude-fit procedure. 
The total systematic uncertainty is obtained by the sum in quadrature of the contributions from different sources, with positive and negative variations treated separately.

In conclusion, an amplitude analysis of the full phase space of \BdToDDKpi decays is performed,
which reveals a new resonance in the \Dp\Kp\pim system, \DsMeson{1}{\val{newDs_Mass_mev}}, with a statistical significance exceeding $10$ standard deviations.
The Breit--Wigner parameters of this state are determined to be
\mbox{$m_0 =
\val{newDs_Mass_mev}\,
^{+\val{newDs_Mass_stat_upper_mev}}_{-\val{newDs_Mass_stat_lower_mev}}\stat\,
^{+\val{newDs_Mass_syst_upper_mev}}_{-\val{newDs_Mass_syst_lower_mev}}\syst
\mev
$}
and
\mbox{$\Gamma_0 =
\val{newDs_Width_mev}\,
^{+\val{newDs_Width_stat_upper_mev}}_{-\val{newDs_Width_stat_lower_mev}}\stat\,
^{+\phantom{0}\val{newDs_Width_syst_upper_mev}}_{-\val{newDs_Width_syst_lower_mev}}\syst
\mev
$},
with its spin-parity found to be $J^P=1^+$. 
Fit fractions of the individual components in the amplitude fit are also measured.
The $\DsMeson{1}{\val{newDs_Mass_mev}}$ state is compatible with a conventional $D_s(2P^{(\prime)}_{1})^+$ state, corresponding to the first radial excitation of the $P$-wave $D_s$ system commonly identified with the \DsMeson{1}{2460}/\DsMeson{1}{2536} states. 
Its discovery provides valuable input for understanding the spectra of the \Ds system and will help clarify the longstanding puzzles related to these low-lying states.

%% file: LHCb/acknowledgements.tex
\section*{Acknowledgements}
%
%
\noindent We express our gratitude to our colleagues in the CERN
accelerator departments for the excellent performance of the LHC. We
thank the technical and administrative staff at the LHCb
institutes.
We acknowledge support from CERN and from the national agencies:
ARC (Australia);
CAPES, CNPq, FAPERJ and FINEP (Brazil); 
MOST and NSFC (China); 
CNRS/IN2P3 and CEA (France);  
BMFTR, DFG and MPG (Germany);
INFN (Italy); 
NWO (Netherlands); 
MNiSW and NCN (Poland); 
MEC/IFA (Romania); 
MICIU and AEI (Spain);
SNSF and SER (Switzerland); 
NASU (Ukraine); 
STFC (United Kingdom); 
DOE NP and NSF (USA).
We acknowledge the computing resources that are provided by ARDC (Australia), 
CBPF (Brazil),
CERN, 
IHEP and LZU (China),
IN2P3 (France), 
KIT and DESY (Germany), 
INFN (Italy), 
SURF (Netherlands),
Polish WLCG (Poland),
IFIN-HH (Romania), 
PIC (Spain), CSCS (Switzerland), 
GridPP (United Kingdom),
and NSF (USA).  
We are indebted to the communities behind the multiple open-source
software packages on which we depend.
Individual groups or members have received support from
RTP (Australia), 
Key Research Program of Frontier Sciences of CAS, CAS PIFI, CAS CCEPP (China); 
Minciencias (Colombia);
EPLANET, Marie Sk\l{}odowska-Curie Actions, ERC and NextGenerationEU (European Union);
A*MIDEX, ANR, IPhU and Labex P2IO, and R\'{e}gion Auvergne-Rh\^{o}ne-Alpes (France);
Alexander-von-Humboldt Foundation (Germany);
ICSC (Italy); 
Severo Ochoa and Mar\'ia de Maeztu Units of Excellence, GVA, XuntaGal, GENCAT, InTalent-Inditex and Prog.~Atracci\'on Talento CM (Spain);
the Leverhulme Trust, the Royal Society and UKRI (United Kingdom).

%% file: supplemental.tex
\clearpage

\section*{End Matter}
\label{sec:Supplemental}

\subsection*{Projections of the amplitude-analysis fit}

\begin{figure}[hb]
    \centering
    \includegraphics[width=0.45\linewidth]{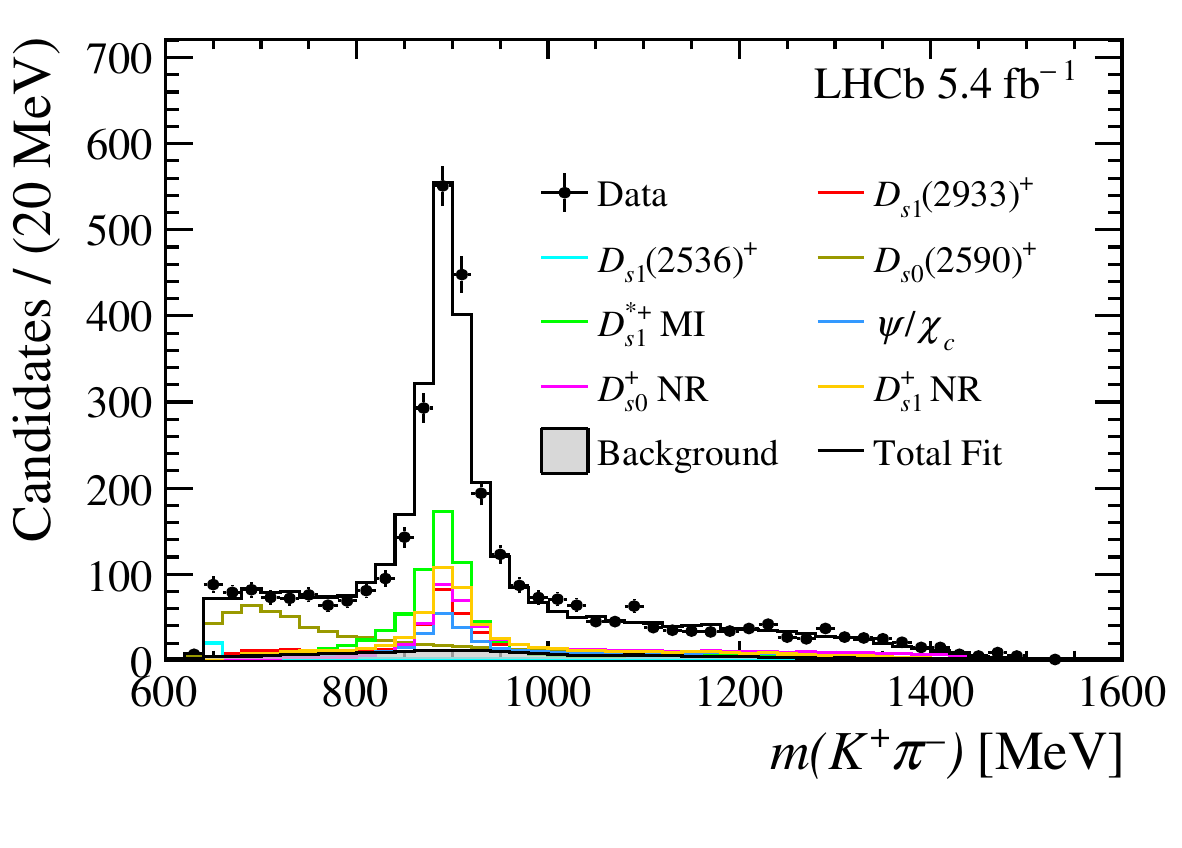}
    \includegraphics[width=0.45\linewidth]{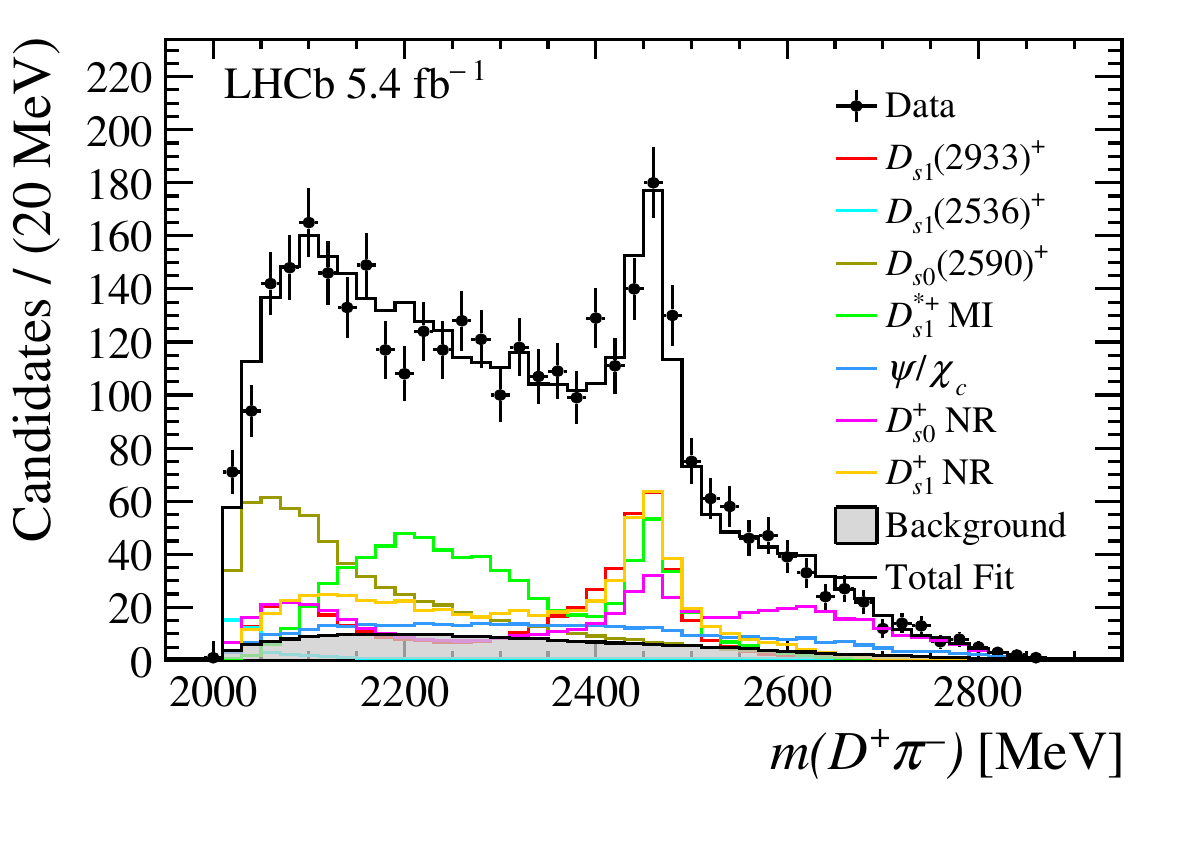}\\
    \includegraphics[width=0.45\linewidth]{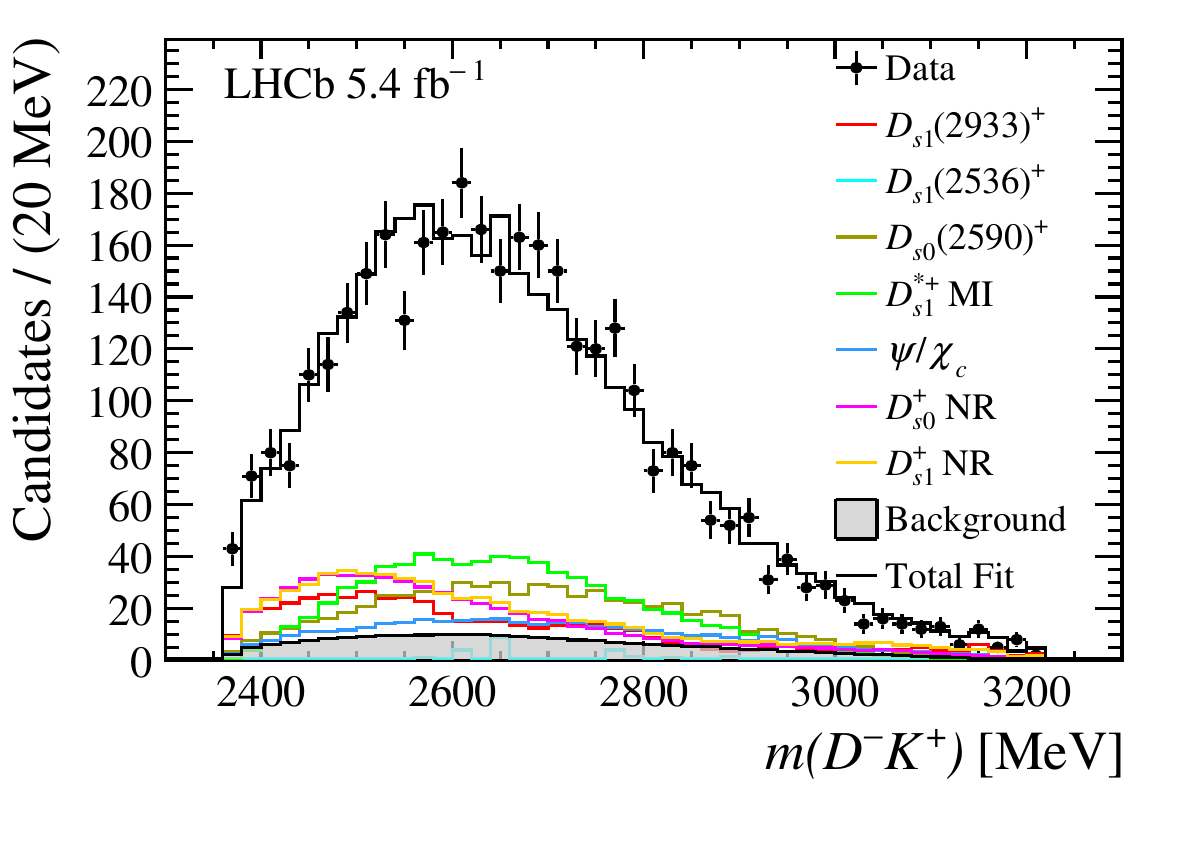}
    \includegraphics[width=0.45\linewidth]{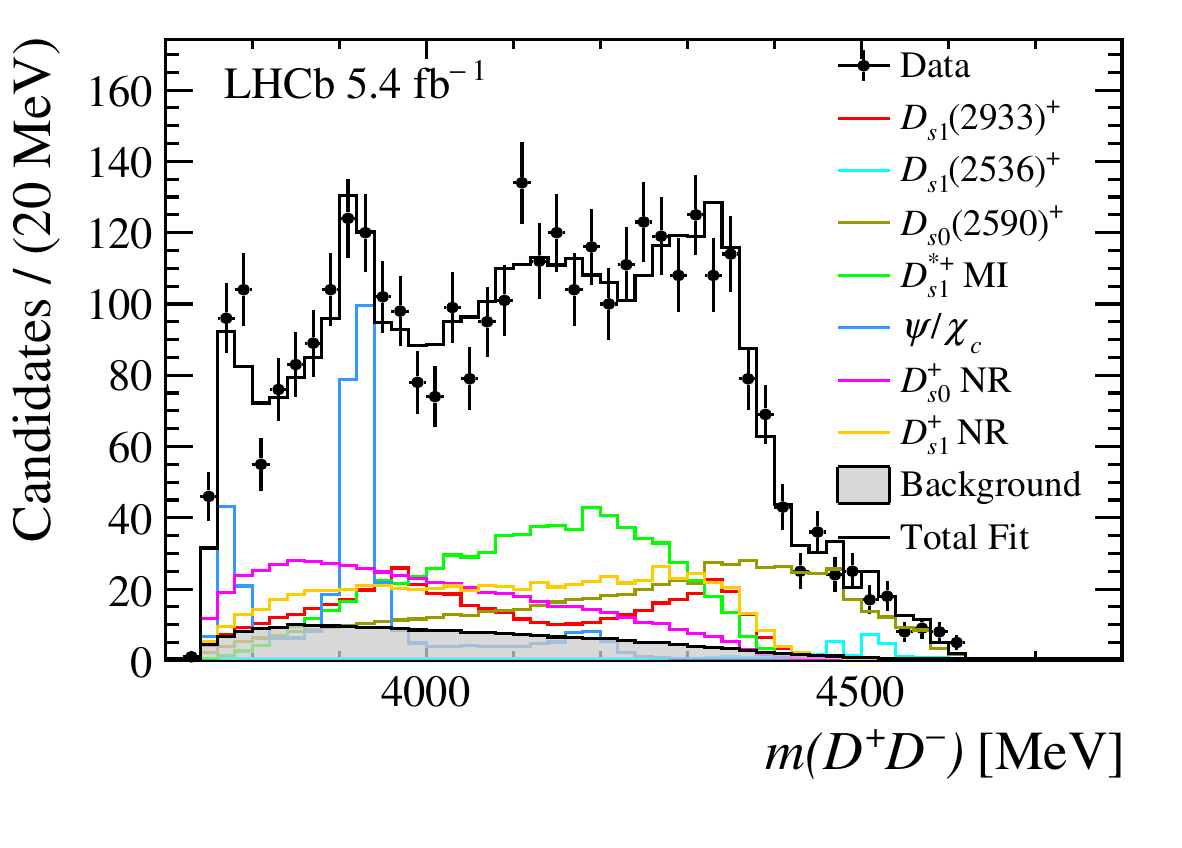}\\
    \includegraphics[width=0.45\linewidth]{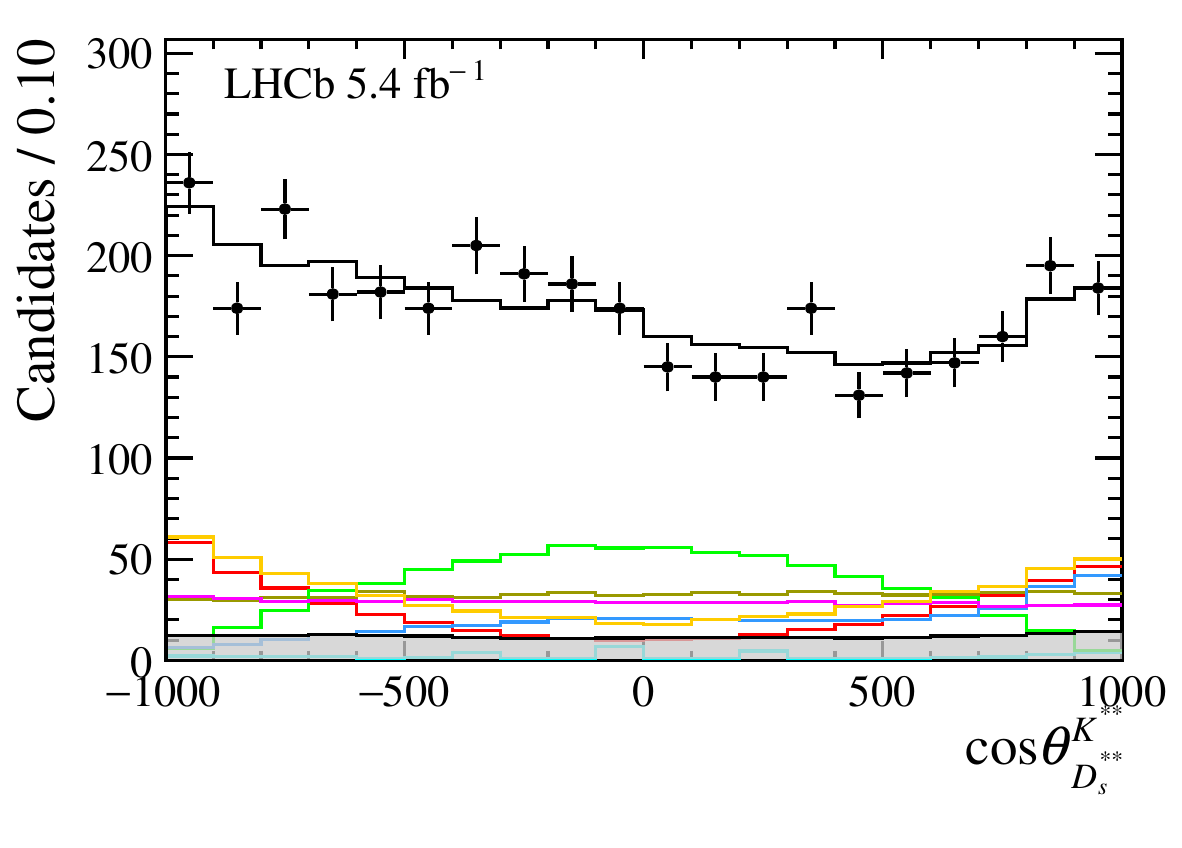}
    \includegraphics[width=0.45\linewidth]{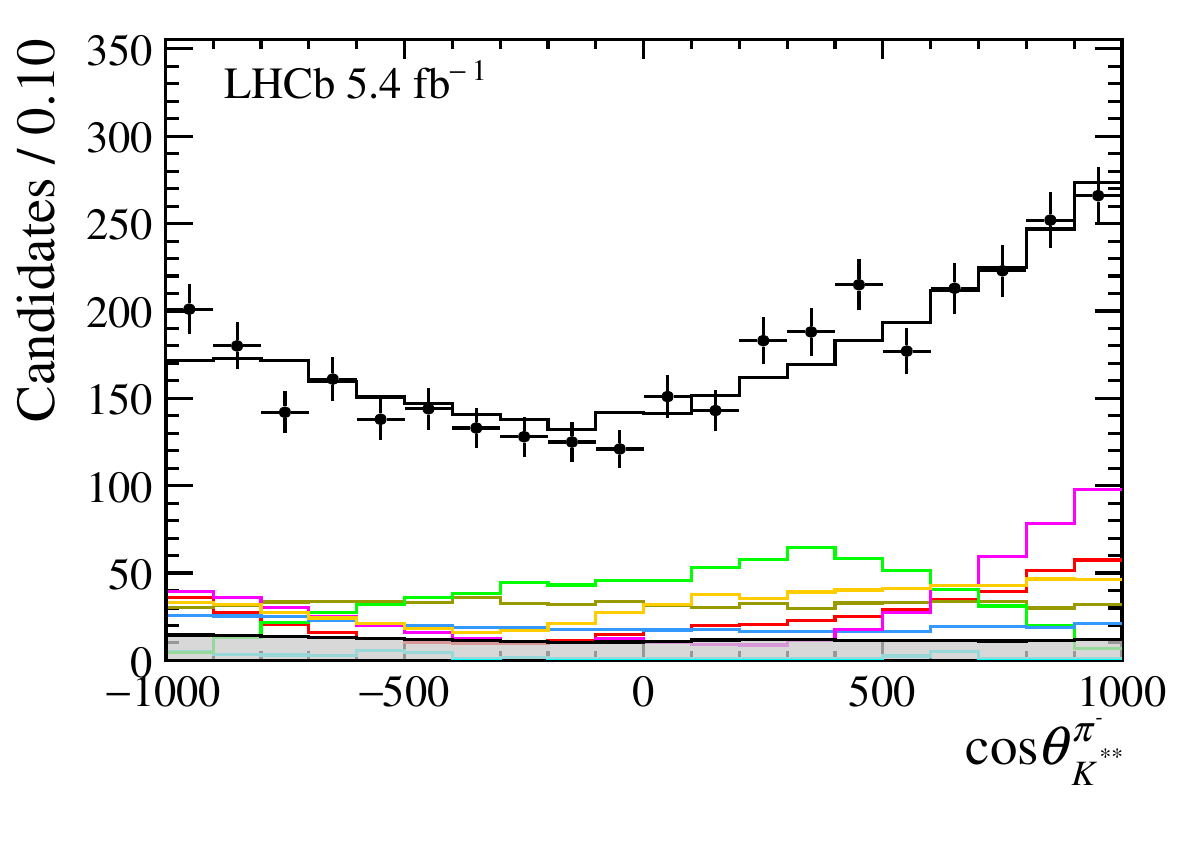}\\
    \includegraphics[width=0.45\linewidth]{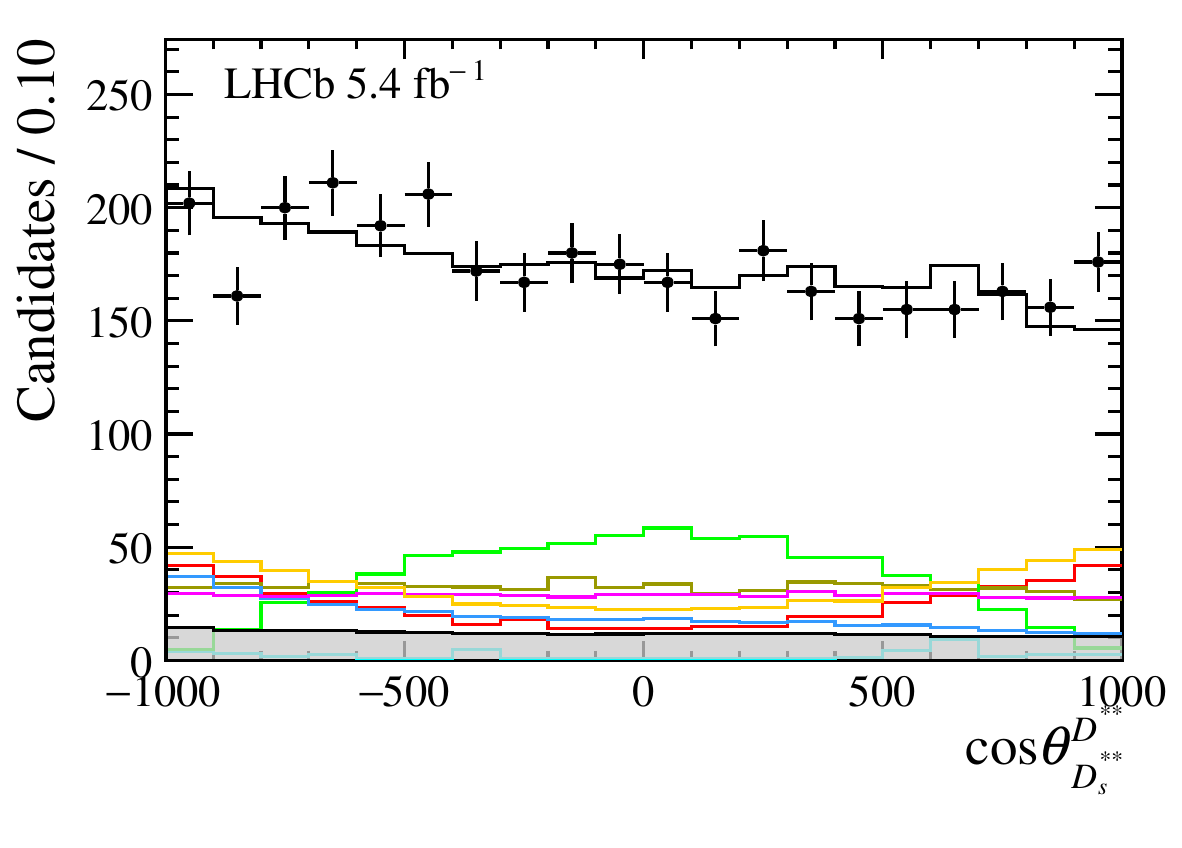}
    \includegraphics[width=0.45\linewidth]{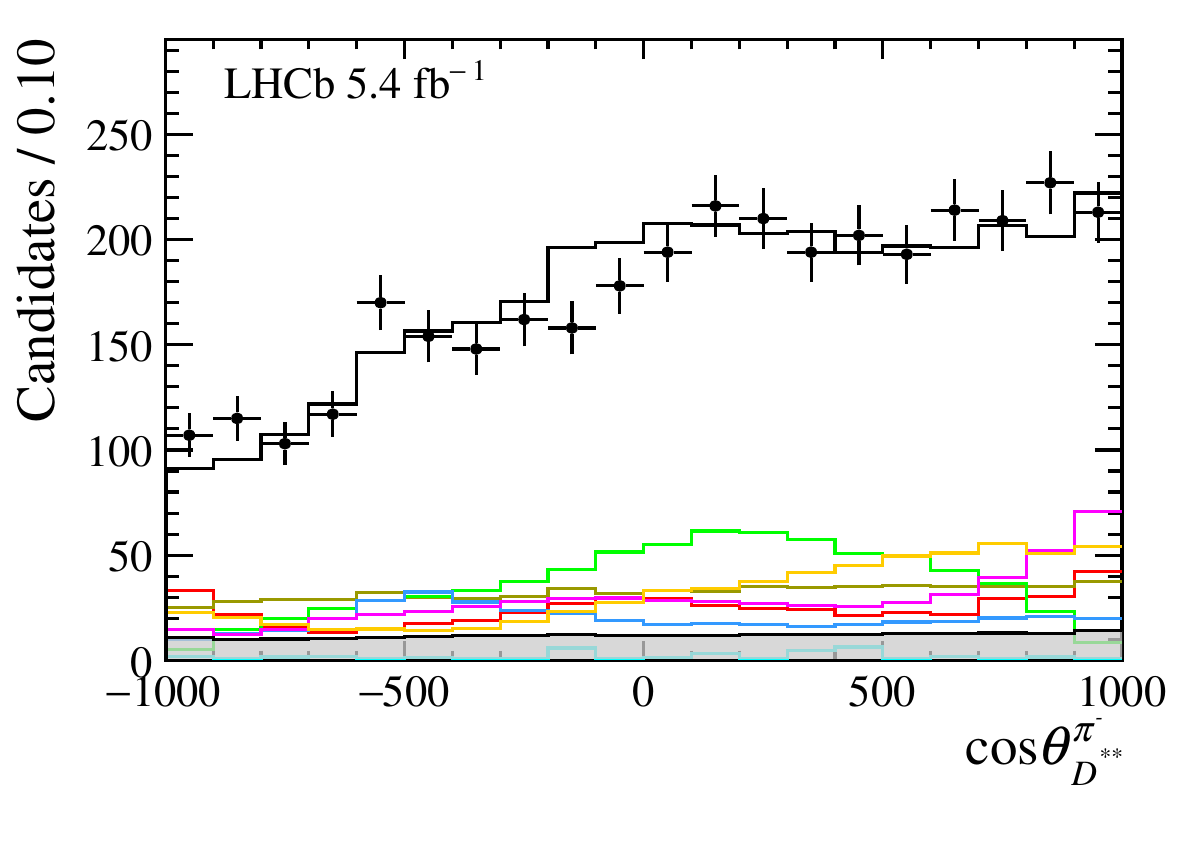}\\
    \caption{Projections for $\BdToDDKpi$ candidates in the signal region with results of the baseline fit. }
    \label{fig:otherProjections}
\end{figure}

Projections of the amplitude-analysis fit for \BdToDDKpi candidates in the signal region in addition to Fig.~\ref{fig:Model_mDKpi} (right) are shown in Fig.~\ref{fig:otherProjections}.

\subsection*{Systematic uncertainties}
The complete sets of systematic uncertainties are listed in Tables~\ref{tab:paras_summary} and \ref{tab:ff_summary} for mass-lineshape parameters and amplitude fit fractions, respectively.

\begin{table}[h]
\centering
\renewcommand{\arraystretch}{1.2}
\caption{Resonance parameters of the $\DsMeson{0}{2590}$ and $\DsMeson{1}{2933}$ states together with uncertainties broken down by origin.}
\label{tab:paras_summary}
\begin{tabular}{lcccc}
\toprule
& \multicolumn{2}{c}{$\DsMeson{0}{2590}$} & \multicolumn{2}{c}{$\DsMeson{1}{2933}$} \\
\cline{2-3} \cline{4-5}
 & $m$ (MeV) & $\Gamma$ (MeV) & $m$ (MeV) & $\Gamma$ (MeV) \\
\midrule
Baseline value & 2606 & 87 & 2933 & 72 \\
Stat & $\asymtight{2}{5}$ & $\asymtight{\phantom{0}6}{13}$ & $\asymtight{6}{5}$ & $\asymtight{18}{11}$ \\
Total syst & $\asymtight{2}{4}$ & $\asymtight{\phantom{0}7}{14}$ & $\asymtight{4}{3}$ & $\asymtight{\phantom{0}7}{10}$ \\
\midrule
Lineshapes & $\asymtight{0}{2}$ & $\asymtight{\phantom{0}4}{12}$ & $\asymtight{0}{2}$ & $\asymtight{5}{6}$ \\
Corrections & $\asymtight{0}{2}$ & $\asymtight{1}{2}$ & $\asymtight{1}{0}$ & $\asymtight{0}{2}$ \\
Parametrizations & $\asymtight{0}{1}$ & $\asymtight{0}{0}$ & $\asymtight{1}{0}$ & $\asymtight{1}{1}$ \\
Fixed parameters & $\asymtight{0}{0}$ & $\asymtight{1}{1}$ & $\asymtight{0}{1}$ & $\asymtight{2}{4}$ \\
Additional resonances & $\asymtight{0}{2}$ & $\asymtight{1}{3}$ & $\asymtight{0}{1}$ & $\asymtight{2}{3}$ \\
Background fraction & $\asymtight{0}{0}$ & $\asymtight{1}{1}$ & $\asymtight{0}{0}$ & $\asymtight{0}{0}$ \\
Background treatment & $\asymtight{0}{2}$ & $\asymtight{0}{6}$ & $\asymtight{2}{0}$ & $\asymtight{4}{5}$ \\
Fit bias & $\asymtight{1}{0}$ & $\asymtight{5}{0}$ & $\asymtight{3}{0}$ & $\asymtight{0}{2}$ \\
\bottomrule
\end{tabular}%
\end{table}

\begin{table}[h]
\renewcommand{\arraystretch}{1.15}
\centering
\caption{Measured fit fractions with uncertainties broken down by origin. The symbol $\mathcal{R}_1$ refers to a \Ds or charmonium (non)resonant contribution, while $\mathcal{R}_2$ refers to a follow-up (non)resonant contribution when $\mathcal{R}_1$ is \Ds, or to an accompanying (non)resonant contribution when $\mathcal{R}_1$ is a charmonium state.}
\label{tab:ff_summary}
\resizebox{\textwidth}{!}{%
\begin{tabular}{llccc|cccccccc}
\toprule
\rotatebox{90}{$\mathcal{R}_1$} & \rotatebox{90}{$\mathcal{R}_2$} & \rotatebox{90}{Fraction (\%)} & \rotatebox{90}{Stat. unc.} & \rotatebox{90}{Total syst. unc.} & \rotatebox{90}{Lineshapes} & \rotatebox{90}{Corrections} & \rotatebox{90}{Parametrizations} & \rotatebox{90}{Fixed parameters} & \rotatebox{90}{Additional resonances} & \rotatebox{90}{Background fraction} & \rotatebox{90}{Background treatment} & \rotatebox{90}{Fit bias} \\
\midrule
\multirow{4}{*}{$\DsMeson{1}{2933}$} & $(\Dp\pim)_S$ & 2.4 & $\asymtight{0.6}{0.7}$ & $\asymtight{0.4}{1.3}$ & $\asymtight{0.4}{1.2}$ & $\asymtight{0.1}{0.0}$ & $\asymtight{0.1}{0.0}$ & $\asymtight{0.0}{0.3}$ & $\asymtight{0.0}{0.1}$ & $\asymtight{0.0}{0.0}$ & $\asymtight{0.1}{0.1}$ & $\asymtight{0.0}{0.5}$ \\
& $\DstzMeson{2}{2460}$ & 7.0 & $\asymtight{6.0}{3.0}$ & $\asymtight{3.0}{3.7}$ & $\asymtight{2.0}{2.3}$ & $\asymtight{0.2}{0.0}$ & $\asymtight{0.2}{0.0}$ & $\asymtight{1.4}{2.8}$ & $\asymtight{1.0}{0.4}$ & $\asymtight{0.1}{0.0}$ & $\asymtight{1.0}{0.8}$ & $\asymtight{0.9}{0.0}$ \\
& $\KstzMeson{1}{892}$ & 8.9 & $\asymtight{1.8}{2.2}$ & $\asymtight{2.4}{1.8}$ & $\asymtight{0.4}{1.0}$ & $\asymtight{0.0}{0.4}$ & $\asymtight{0.1}{0.5}$ & $\asymtight{2.3}{1.1}$ & $\asymtight{0.4}{0.5}$ & $\asymtight{0.0}{0.1}$ & $\asymtight{0.1}{0.7}$ & $\asymtight{0.1}{0.0}$ \\
\cline{2-13}
& Total & 15.6 & $\asymtight{7.4}{3.5}$ & $\asymtight{5.0}{4.9}$ & $\asymtight{3.6}{2.2}$ & $\asymtight{0.2}{0.3}$ & $\asymtight{0.3}{0.3}$ & $\asymtight{3.0}{4.0}$ & $\asymtight{1.3}{0.5}$ & $\asymtight{0.1}{0.1}$ & $\asymtight{0.9}{1.5}$ & $\asymtight{0.3}{0.0}$ \\
\midrule
\multirow{1}{*}{$\DsMeson{0}{2590}$} & $(\Dp\pim)_S$ & 21.5 & $\asymtight{0.5}{2.1}$ & $\asymtight{2.5}{0.6}$ & $\asymtight{2.3}{0.3}$ & $\asymtight{0.1}{0.0}$ & $\asymtight{0.2}{0.0}$ & $\asymtight{0.1}{0.0}$ & $\asymtight{0.9}{0.3}$ & $\asymtight{0.1}{0.0}$ & $\asymtight{0.2}{0.4}$ & $\asymtight{0.0}{0.2}$ \\
\midrule
\multirow{1}{*}{$\DsMeson{1}{2536}$} & $(\Dp\pim)_S$ & 1.0 & $\asymtight{0.1}{0.3}$ & $\asymtight{0.6}{0.4}$ & $\asymtight{0.6}{0.2}$ & $\asymtight{0.0}{0.0}$ & $\asymtight{0.0}{0.0}$ & $\asymtight{0.0}{0.2}$ & $\asymtight{0.0}{0.0}$ & $\asymtight{0.0}{0.0}$ & $\asymtight{0.1}{0.1}$ & $\asymtight{0.0}{0.2}$ \\
\midrule
\multirow{6}{*}{$\DsMesonNoMass{0}\,\text{NR}$} & $\DstzMeson{1}{2600}$ & 1.5 & $\asymtight{0.8}{0.3}$ & $\asymtight{0.4}{0.4}$ & $\asymtight{0.3}{0.3}$ & $\asymtight{0.1}{0.0}$ & $\asymtight{0.1}{0.0}$ & $\asymtight{0.0}{0.3}$ & $\asymtight{0.0}{0.1}$ & $\asymtight{0.0}{0.0}$ & $\asymtight{0.1}{0.1}$ & $\asymtight{0.2}{0.0}$ \\
& $\DstzMeson{2}{2460}$ & 1.9 & $\asymtight{0.5}{0.6}$ & $\asymtight{0.3}{0.9}$ & $\asymtight{0.0}{0.9}$ & $\asymtight{0.1}{0.0}$ & $\asymtight{0.1}{0.0}$ & $\asymtight{0.1}{0.2}$ & $\asymtight{0.1}{0.2}$ & $\asymtight{0.0}{0.0}$ & $\asymtight{0.1}{0.0}$ & $\asymtight{0.0}{0.0}$ \\
& $\KstzMeson{1}{892}$ & 8.4 & $\asymtight{2.2}{0.8}$ & $\asymtight{1.9}{0.9}$ & $\asymtight{1.1}{0.5}$ & $\asymtight{0.2}{0.0}$ & $\asymtight{0.3}{0.0}$ & $\asymtight{0.6}{0.0}$ & $\asymtight{1.2}{0.3}$ & $\asymtight{0.0}{0.0}$ & $\asymtight{0.2}{0.7}$ & $\asymtight{0.7}{0.0}$ \\
& $\KstzMeson{1}{1410}$ & 1.6 & $\asymtight{1.1}{0.3}$ & $\asymtight{0.7}{0.9}$ & $\asymtight{0.0}{0.5}$ & $\asymtight{0.1}{0.0}$ & $\asymtight{0.1}{0.0}$ & $\asymtight{0.0}{0.3}$ & $\asymtight{0.0}{0.7}$ & $\asymtight{0.0}{0.0}$ & $\asymtight{0.6}{0.0}$ & $\asymtight{0.0}{0.0}$ \\
& $\KstzMeson{2}{1430}$ & 1.0 & $\asymtight{0.5}{0.3}$ & $\asymtight{0.9}{0.1}$ & $\asymtight{0.3}{0.0}$ & $\asymtight{0.1}{0.0}$ & $\asymtight{0.1}{0.0}$ & $\asymtight{0.5}{0.0}$ & $\asymtight{0.7}{0.0}$ & $\asymtight{0.0}{0.0}$ & $\asymtight{0.1}{0.0}$ & $\asymtight{0.0}{0.1}$ \\
\cline{2-13}
& Total & 17.3 & $\asymtight{3.2}{0.1}$ & $\asymtight{2.1}{0.7}$ & $\asymtight{0.1}{0.6}$ & $\asymtight{0.9}{0.0}$ & $\asymtight{1.0}{0.0}$ & $\asymtight{0.2}{0.0}$ & $\asymtight{0.5}{0.3}$ & $\asymtight{0.0}{0.0}$ & $\asymtight{1.2}{0.0}$ & $\asymtight{0.7}{0.0}$ \\
\midrule
\multirow{4}{*}{$\DsMesonNoMass{1}\,\text{NR}$} & $\DstzMeson{1}{2600}$ & 2.5 & $\asymtight{0.6}{0.5}$ & $\asymtight{0.6}{0.4}$ & $\asymtight{0.5}{0.1}$ & $\asymtight{0.0}{0.1}$ & $\asymtight{0.0}{0.1}$ & $\asymtight{0.0}{0.3}$ & $\asymtight{0.3}{0.0}$ & $\asymtight{0.0}{0.0}$ & $\asymtight{0.0}{0.1}$ & $\asymtight{0.1}{0.0}$ \\
& $\DstzMeson{2}{2460}$ & 6.0 & $\asymtight{7.0}{3.0}$ & $\asymtight{2.6}{4.4}$ & $\asymtight{2.0}{3.1}$ & $\asymtight{0.0}{0.0}$ & $\asymtight{0.5}{0.0}$ & $\asymtight{0.0}{2.6}$ & $\asymtight{1.0}{1.4}$ & $\asymtight{0.0}{0.0}$ & $\asymtight{1.0}{1.0}$ & $\asymtight{0.7}{0.0}$ \\
& $\KstzMeson{1}{892}$ & 10.0 & $\asymtight{2.3}{1.1}$ & $\asymtight{1.8}{1.2}$ & $\asymtight{0.5}{0.8}$ & $\asymtight{0.1}{0.0}$ & $\asymtight{0.2}{0.0}$ & $\asymtight{0.9}{0.9}$ & $\asymtight{1.4}{0.2}$ & $\asymtight{0.0}{0.0}$ & $\asymtight{0.2}{0.2}$ & $\asymtight{0.6}{0.0}$ \\
\cline{2-13}
& Total & 18.9 & $\asymtight{7.3}{2.5}$ & $\asymtight{3.3}{4.5}$ & $\asymtight{1.3}{2.5}$ & $\asymtight{0.1}{0.1}$ & $\asymtight{0.5}{0.0}$ & $\asymtight{0.4}{3.4}$ & $\asymtight{2.1}{1.0}$ & $\asymtight{0.0}{0.0}$ & $\asymtight{1.4}{1.0}$ & $\asymtight{1.5}{0.0}$ \\
\midrule
\multirow{3}{*}{$\DsstMesonNoMass{1}\,\text{MI}$} & $\DstzMeson{2}{2460}$ & 4.7 & $\asymtight{0.8}{0.5}$ & $\asymtight{0.4}{0.3}$ & $\asymtight{0.2}{0.0}$ & $\asymtight{0.0}{0.0}$ & $\asymtight{0.0}{0.1}$ & $\asymtight{0.1}{0.1}$ & $\asymtight{0.1}{0.0}$ & $\asymtight{0.0}{0.0}$ & $\asymtight{0.3}{0.3}$ & $\asymtight{0.1}{0.0}$ \\
& $\KstzMeson{1}{892}$ & 18.0 & $\asymtight{0.9}{2.4}$ & $\asymtight{0.4}{2.0}$ & $\asymtight{0.2}{1.4}$ & $\asymtight{0.0}{0.5}$ & $\asymtight{0.1}{0.1}$ & $\asymtight{0.2}{0.2}$ & $\asymtight{0.1}{1.1}$ & $\asymtight{0.0}{0.1}$ & $\asymtight{0.3}{0.4}$ & $\asymtight{0.0}{0.5}$ \\
\cline{2-13}
& Total & 22.2 & $\asymtight{1.5}{2.1}$ & $\asymtight{0.9}{1.8}$ & $\asymtight{0.2}{1.4}$ & $\asymtight{0.1}{0.6}$ & $\asymtight{0.1}{0.1}$ & $\asymtight{0.0}{0.1}$ & $\asymtight{0.3}{0.8}$ & $\asymtight{0.0}{0.2}$ & $\asymtight{0.8}{0.4}$ & $\asymtight{0.0}{0.3}$ \\
\midrule
\multirow{2}{*}{$\chiczero(3915)$} & $\KstzMeson{1}{892}$ & 0.9 & $\asymtight{0.5}{0.3}$ & $\asymtight{0.3}{0.1}$ & $\asymtight{0.2}{0.0}$ & $\asymtight{0.0}{0.0}$ & $\asymtight{0.0}{0.0}$ & $\asymtight{0.1}{0.0}$ & $\asymtight{0.0}{0.1}$ & $\asymtight{0.0}{0.0}$ & $\asymtight{0.2}{0.1}$ & $\asymtight{0.0}{0.0}$ \\
& $(\Kp\pim)_S$ & 6.7 & $\asymtight{0.1}{2.1}$ & $\asymtight{0.4}{1.1}$ & $\asymtight{0.0}{0.2}$ & $\asymtight{0.0}{0.6}$ & $\asymtight{0.1}{0.5}$ & $\asymtight{0.4}{0.0}$ & $\asymtight{0.0}{0.3}$ & $\asymtight{0.0}{0.1}$ & $\asymtight{0.2}{0.7}$ & $\asymtight{0.0}{0.2}$ \\
\midrule
\multirow{1}{*}{$\chictwo(3930)$} & $\KstzMeson{1}{892}$ & 0.7 & $\asymtight{0.4}{0.3}$ & $\asymtight{0.2}{0.2}$ & $\asymtight{0.1}{0.1}$ & $\asymtight{0.0}{0.1}$ & $\asymtight{0.0}{0.1}$ & $\asymtight{0.0}{0.0}$ & $\asymtight{0.1}{0.0}$ & $\asymtight{0.0}{0.0}$ & $\asymtight{0.2}{0.1}$ & $\asymtight{0.0}{0.0}$ \\
\midrule
\multirow{2}{*}{$\psi(3770)$} & $\KstzMeson{1}{892}$ & 2.3 & $\asymtight{0.5}{0.4}$ & $\asymtight{0.4}{0.5}$ & $\asymtight{0.0}{0.2}$ & $\asymtight{0.0}{0.0}$ & $\asymtight{0.1}{0.0}$ & $\asymtight{0.3}{0.1}$ & $\asymtight{0.0}{0.4}$ & $\asymtight{0.0}{0.0}$ & $\asymtight{0.3}{0.1}$ & $\asymtight{0.1}{0.0}$ \\
& $\KstzMeson{2}{1430}$ & 0.9 & $\asymtight{0.5}{0.3}$ & $\asymtight{0.0}{0.3}$ & $\asymtight{0.0}{0.1}$ & $\asymtight{0.0}{0.1}$ & $\asymtight{0.0}{0.1}$ & $\asymtight{0.0}{0.0}$ & $\asymtight{0.0}{0.1}$ & $\asymtight{0.0}{0.0}$ & $\asymtight{0.0}{0.2}$ & $\asymtight{0.0}{0.0}$ \\
\midrule
\multirow{1}{*}{$\psi(4040)$} & $\KstzMeson{1}{892}$ & 0.9 & $\asymtight{0.6}{0.3}$ & $\asymtight{0.7}{0.3}$ & $\asymtight{0.2}{0.0}$ & $\asymtight{0.0}{0.2}$ & $\asymtight{0.0}{0.2}$ & $\asymtight{0.1}{0.0}$ & $\asymtight{0.6}{0.1}$ & $\asymtight{0.0}{0.0}$ & $\asymtight{0.0}{0.1}$ & $\asymtight{0.2}{0.0}$ \\
\midrule
\multirow{1}{*}{$\psi(4160)$} & $\KstzMeson{1}{892}$ & 0.9 & $\asymtight{0.7}{0.1}$ & $\asymtight{1.8}{0.3}$ & $\asymtight{0.0}{0.2}$ & $\asymtight{0.2}{0.0}$ & $\asymtight{0.2}{0.0}$ & $\asymtight{0.0}{0.0}$ & $\asymtight{1.8}{0.2}$ & $\asymtight{0.0}{0.0}$ & $\asymtight{0.1}{0.0}$ & $\asymtight{0.1}{0.0}$ \\
\midrule
\multirow{1}{*}{Total} & Total & 109.7 & $\asymtight{12.4}{\phantom{0}5.4}$ & $\asymtight{7.5}{9.5}$ & $\asymtight{4.1}{4.2}$ & $\asymtight{0.0}{0.4}$ & $\asymtight{0.7}{0.4}$ & $\asymtight{5.0}{6.9}$ & $\asymtight{2.1}{1.9}$ & $\asymtight{0.0}{0.1}$ & $\asymtight{2.5}{4.6}$ & $\asymtight{1.9}{0.0}$ \\
\bottomrule
\end{tabular}
}
\end{table}

\clearpage

%% file: Authorship_LHCb-PAPER-2025-073.tex
\centerline
{\large\bf LHCb collaboration}
\begin
{flushleft}
\small
R.~Aaij$^{38}$\lhcborcid{0000-0003-0533-1952},
M. ~Abdelfatah$^{69}$,
A.S.W.~Abdelmotteleb$^{57}$\lhcborcid{0000-0001-7905-0542},
C.~Abellan~Beteta$^{51}$\lhcborcid{0009-0009-0869-6798},
F.~Abudin\'en$^{59}$\lhcborcid{0000-0002-6737-3528},
T.~Ackernley$^{61}$\lhcborcid{0000-0002-5951-3498},
A. A. ~Adefisoye$^{69}$\lhcborcid{0000-0003-2448-1550},
B.~Adeva$^{47}$\lhcborcid{0000-0001-9756-3712},
M.~Adinolfi$^{55}$\lhcborcid{0000-0002-1326-1264},
P.~Adlarson$^{87}$\lhcborcid{0000-0001-6280-3851}
% ...

{\footnotesize \it

% ...
$^{38}$Nikhef National Institute for Subatomic Physics, Amsterdam, Netherlands\\
% ...
$^{47}$Instituto Galego de F{\'\i}sica de Altas Enerx{\'\i}as (IGFAE), Universidade de Santiago de Compostela, Santiago de Compostela, Spain\\
% ...
$^{51}$Physik-Institut, Universit{\"a}t Z{\"u}rich, Z{\"u}rich, Switzerland\\
% ...
$^{55}$H.H. Wills Physics Laboratory, University of Bristol, Bristol, United Kingdom\\
% ...
$^{57}$Department of Physics, University of Warwick, Coventry, United Kingdom\\
% ...
$^{59}$School of Physics and Astronomy, University of Edinburgh, Edinburgh, United Kingdom\\
$^{60}$School of Physics and Astronomy, University of Glasgow, Glasgow, United Kingdom\\
$^{61}$Oliver Lodge Laboratory, University of Liverpool, Liverpool, United Kingdom\\
% ...
$^{69}$Syracuse University, Syracuse, NY, United States\\
% ...
$^{87}$Department of Physics and Astronomy, Uppsala University, Uppsala, Sweden, associated to $^{60}$\\
% ...
}
\end{flushleft}

%% file: main.bbl
\ifx\mcitethebibliography\mciteundefinedmacro
\PackageError{LHCb.bst}{mciteplus.sty has not been loaded}
{This bibstyle requires the use of the mciteplus package.}\fi
\providecommand{\href}[2]{#2}
\begin{mcitethebibliography}{10}
\mciteSetBstSublistMode{n}
\mciteSetBstMaxWidthForm{subitem}{\alph{mcitesubitemcount})}
\mciteSetBstSublistLabelBeginEnd{\mcitemaxwidthsubitemform\space}
{\relax}{\relax}

\bibitem{Guo2019CharmSpec}
F.~K. Guo, \ifthenelse{\boolean{articletitles}}{\emph{Status of charmed meson spectroscopy}, }{}\href{https://doi.org/10.1051/epjconf/201920202001}{EPJ Web Conf.\  \textbf{202} (2019) 02001}, Proceedings of CHARM 2018\relax
\mciteBstWouldAddEndPuncttrue
\mciteSetBstMidEndSepPunct{\mcitedefaultmidpunct}
{\mcitedefaultendpunct}{\mcitedefaultseppunct}\relax
\EndOfBibitem
\bibitem{vanBeveren2021Review}
E.~van Beveren and G.~Rupp, \ifthenelse{\boolean{articletitles}}{\emph{{Modern meson spectroscopy: the fundamental role of unitarity}}, }{}\href{https://doi.org/10.1016/j.ppnp.2020.103845}{Prog.\ Part.\ Nucl.\ Phys.\  \textbf{117} (2021) 103845}, \href{http://arxiv.org/abs/2012.03693}{{\normalfont\ttfamily arXiv:2012.03693}}\relax
\mciteBstWouldAddEndPuncttrue
\mciteSetBstMidEndSepPunct{\mcitedefaultmidpunct}
{\mcitedefaultendpunct}{\mcitedefaultseppunct}\relax
\EndOfBibitem
\bibitem{Godfrey2015CharmSpec}
S.~Godfrey and K.~Moats, \ifthenelse{\boolean{articletitles}}{\emph{{Properties of excited charm and charm-strange mesons}}, }{}\href{https://doi.org/10.1103/PhysRevD.93.034035}{Phys.\ Rev.\  \textbf{D93} (2016) 034035}, \href{http://arxiv.org/abs/1510.08305}{{\normalfont\ttfamily arXiv:1510.08305}}\relax
\mciteBstWouldAddEndPuncttrue
\mciteSetBstMidEndSepPunct{\mcitedefaultmidpunct}
{\mcitedefaultendpunct}{\mcitedefaultseppunct}\relax
\EndOfBibitem
\bibitem{BaBar2003Ds2317}
BaBar collaboration, B.~Aubert {\em et~al.}, \ifthenelse{\boolean{articletitles}}{\emph{{Observation of a narrow meson decaying to $D_s^+ \pi^0$ at a mass of 2.32 GeV/c$^2$}}, }{}\href{https://doi.org/10.1103/PhysRevLett.90.242001}{Phys.\ Rev.\ Lett.\  \textbf{90} (2003) 242001}, \href{http://arxiv.org/abs/hep-ex/0304021}{{\normalfont\ttfamily arXiv:hep-ex/0304021}}\relax
\mciteBstWouldAddEndPuncttrue
\mciteSetBstMidEndSepPunct{\mcitedefaultmidpunct}
{\mcitedefaultendpunct}{\mcitedefaultseppunct}\relax
\EndOfBibitem
\bibitem{CLEO:2003ggt}
CLEO collaboration, D.~Besson {\em et~al.}, \ifthenelse{\boolean{articletitles}}{\emph{{Observation of a narrow resonance of mass $2.46\gevcc$ decaying to $D_s^{*+} \piz$ and confirmation of the $D^*_{sJ}(2317)$ state}}, }{}\href{https://doi.org/10.1103/PhysRevD.68.032002}{Phys.\ Rev.\  \textbf{D68} (2003) 032002}, Erratum \href{https://doi.org/10.1103/PhysRevD.75.119908}{ibid.\   \textbf{D75} (2007) 119908}, \href{http://arxiv.org/abs/hep-ex/0305100}{{\normalfont\ttfamily arXiv:hep-ex/0305100}}\relax
\mciteBstWouldAddEndPuncttrue
\mciteSetBstMidEndSepPunct{\mcitedefaultmidpunct}
{\mcitedefaultendpunct}{\mcitedefaultseppunct}\relax
\EndOfBibitem
\bibitem{BaBar:2003cdx}
BaBar collaboration, B.~Aubert {\em et~al.}, \ifthenelse{\boolean{articletitles}}{\emph{{Observation of a narrow meson decaying to $\Dsp \piz \gamma$ at a mass of $2.458\gevcc$}}, }{}\href{https://doi.org/10.1103/PhysRevD.69.031101}{Phys.\ Rev.\  \textbf{D69} (2004) 031101}, \href{http://arxiv.org/abs/hep-ex/0310050}{{\normalfont\ttfamily arXiv:hep-ex/0310050}}\relax
\mciteBstWouldAddEndPuncttrue
\mciteSetBstMidEndSepPunct{\mcitedefaultmidpunct}
{\mcitedefaultendpunct}{\mcitedefaultseppunct}\relax
\EndOfBibitem
\bibitem{Belle2003Ds2460}
Belle collaboration, P.~Krokovny {\em et~al.}, \ifthenelse{\boolean{articletitles}}{\emph{{Observation of the $D_{sJ}(2317)$ and $D_{sJ}(2457)$ in $B$ decays}}, }{}\href{https://doi.org/10.1103/PhysRevLett.91.262002}{Phys.\ Rev.\ Lett.\  \textbf{91} (2003) 262002}, \href{http://arxiv.org/abs/hep-ex/0308019}{{\normalfont\ttfamily arXiv:hep-ex/0308019}}\relax
\mciteBstWouldAddEndPuncttrue
\mciteSetBstMidEndSepPunct{\mcitedefaultmidpunct}
{\mcitedefaultendpunct}{\mcitedefaultseppunct}\relax
\EndOfBibitem
\bibitem{Mohler2013Ds2317LQCD}
D.~Mohler {\em et~al.}, \ifthenelse{\boolean{articletitles}}{\emph{{$D_{s0}^*(2317)$ meson and $D$-meson-kaon scattering from lattice QCD}}, }{}\href{https://doi.org/10.1103/PhysRevLett.111.222001}{Phys.\ Rev.\ Lett.\  \textbf{111} (2013) 222001}, \href{http://arxiv.org/abs/1308.3175}{{\normalfont\ttfamily arXiv:1308.3175}}\relax
\mciteBstWouldAddEndPuncttrue
\mciteSetBstMidEndSepPunct{\mcitedefaultmidpunct}
{\mcitedefaultendpunct}{\mcitedefaultseppunct}\relax
\EndOfBibitem
\bibitem{Bali2017DsLQCD}
G.~S. Bali, S.~Collins, A.~Cox, and A.~Sch{\"a}fer, \ifthenelse{\boolean{articletitles}}{\emph{{Masses and decay constants of the $D_{s0}^*(2317)$ and $D_{s1}(2460)$ from $N_f=2$ lattice QCD close to the physical point}}, }{}\href{https://doi.org/10.1103/PhysRevD.96.074501}{Phys.\ Rev.\  \textbf{D96} (2017) 074501}, \href{http://arxiv.org/abs/1706.01247}{{\normalfont\ttfamily arXiv:1706.01247}}\relax
\mciteBstWouldAddEndPuncttrue
\mciteSetBstMidEndSepPunct{\mcitedefaultmidpunct}
{\mcitedefaultendpunct}{\mcitedefaultseppunct}\relax
\EndOfBibitem
\bibitem{Ni2022DsSpectrum}
R.-H. Ni, Q.~Li, and X.-H. Zhong, \ifthenelse{\boolean{articletitles}}{\emph{{Mass spectra and strong decays of charmed and charmed-strange mesons}}, }{}\href{https://doi.org/10.1103/PhysRevD.105.056006}{Phys.\ Rev.\  \textbf{D105} (2022) 056006}, \href{http://arxiv.org/abs/2110.05024}{{\normalfont\ttfamily arXiv:2110.05024}}\relax
\mciteBstWouldAddEndPuncttrue
\mciteSetBstMidEndSepPunct{\mcitedefaultmidpunct}
{\mcitedefaultendpunct}{\mcitedefaultseppunct}\relax
\EndOfBibitem
\bibitem{Ni2023UnquenchedModel}
R.-H. Ni, J.-J. Wu, and X.-H. Zhong, \ifthenelse{\boolean{articletitles}}{\emph{{Unified unquenched quark model for heavy-light mesons with chiral dynamics}}, }{}\href{https://doi.org/10.1103/PhysRevD.109.116006}{Phys.\ Rev.\  \textbf{D109} (2024) 116006}, \href{http://arxiv.org/abs/2312.04765}{{\normalfont\ttfamily arXiv:2312.04765}}\relax
\mciteBstWouldAddEndPuncttrue
\mciteSetBstMidEndSepPunct{\mcitedefaultmidpunct}
{\mcitedefaultendpunct}{\mcitedefaultseppunct}\relax
\EndOfBibitem
\bibitem{Barnes:2003dj}
T.~Barnes, F.~E. Close, and H.~J. Lipkin, \ifthenelse{\boolean{articletitles}}{\emph{{Implications of a DK molecule at 2.32 GeV}}, }{}\href{https://doi.org/10.1103/PhysRevD.68.054006}{Phys.\ Rev.\  \textbf{D68} (2003) 054006}, \href{http://arxiv.org/abs/hep-ph/0305025}{{\normalfont\ttfamily arXiv:hep-ph/0305025}}\relax
\mciteBstWouldAddEndPuncttrue
\mciteSetBstMidEndSepPunct{\mcitedefaultmidpunct}
{\mcitedefaultendpunct}{\mcitedefaultseppunct}\relax
\EndOfBibitem
\bibitem{Cheng:2003kg}
H.-Y. Cheng and W.-S. Hou, \ifthenelse{\boolean{articletitles}}{\emph{{B decays as spectroscope for charmed four quark states}}, }{}\href{https://doi.org/10.1016/S0370-2693(03)00834-7}{Phys.\ Lett.\  \textbf{B566} (2003) 193}, \href{http://arxiv.org/abs/hep-ph/0305038}{{\normalfont\ttfamily arXiv:hep-ph/0305038}}\relax
\mciteBstWouldAddEndPuncttrue
\mciteSetBstMidEndSepPunct{\mcitedefaultmidpunct}
{\mcitedefaultendpunct}{\mcitedefaultseppunct}\relax
\EndOfBibitem
\bibitem{Maiani:2004vq}
L.~Maiani, F.~Piccinini, A.~D. Polosa, and V.~Riquer, \ifthenelse{\boolean{articletitles}}{\emph{{Diquark-antidiquarks with hidden or open charm and the nature of X(3872)}}, }{}\href{https://doi.org/10.1103/PhysRevD.71.014028}{Phys.\ Rev.\  \textbf{D71} (2005) 014028}, \href{http://arxiv.org/abs/hep-ph/0412098}{{\normalfont\ttfamily arXiv:hep-ph/0412098}}\relax
\mciteBstWouldAddEndPuncttrue
\mciteSetBstMidEndSepPunct{\mcitedefaultmidpunct}
{\mcitedefaultendpunct}{\mcitedefaultseppunct}\relax
\EndOfBibitem
\bibitem{LHCb-PAPER-2022-026}
LHCb collaboration, R.~Aaij {\em et~al.}, \ifthenelse{\boolean{articletitles}}{\emph{{First observation of a doubly charged tetraquark candidate and its neutral partner}}, }{}\href{https://doi.org/10.1103/PhysRevLett.131.041902}{Phys.\ Rev.\ Lett.\  \textbf{131} (2023) 041902}, \href{http://arxiv.org/abs/2212.02716}{{\normalfont\ttfamily arXiv:2212.02716}}\relax
\mciteBstWouldAddEndPuncttrue
\mciteSetBstMidEndSepPunct{\mcitedefaultmidpunct}
{\mcitedefaultendpunct}{\mcitedefaultseppunct}\relax
\EndOfBibitem
\bibitem{LHCb-PAPER-2022-027}
LHCb collaboration, R.~Aaij {\em et~al.}, \ifthenelse{\boolean{articletitles}}{\emph{{Amplitude analysis of $\Bz \rightarrow \Dzb \Dsp \pim$ and $\Bp \rightarrow \Dm \Dsp\pip$ decays}}, }{}\href{https://doi.org/10.1103/PhysRevD.108.012017}{Phys.\ Rev.\  \textbf{D108} (2023) 012017}, \href{http://arxiv.org/abs/2212.02717}{{\normalfont\ttfamily arXiv:2212.02717}}\relax
\mciteBstWouldAddEndPuncttrue
\mciteSetBstMidEndSepPunct{\mcitedefaultmidpunct}
{\mcitedefaultendpunct}{\mcitedefaultseppunct}\relax
\EndOfBibitem
\bibitem{LHCb-PAPER-2024-033}
LHCb collaboration, R.~Aaij {\em et~al.}, \ifthenelse{\boolean{articletitles}}{\emph{{Study of $D_{s1}(2460)^+ \to D_s^+ \pi^+ \pi^-$ in $B \to \Dbar^{(*)} D_s^+ \pi^+ \pi^-$ decays}}, }{}\href{https://doi.org/10.1016/j.scib.2025.02.025}{{Science Bulletin} \textbf{70} (2025) 1432}, \href{http://arxiv.org/abs/2411.03399}{{\normalfont\ttfamily arXiv:2411.03399}}\relax
\mciteBstWouldAddEndPuncttrue
\mciteSetBstMidEndSepPunct{\mcitedefaultmidpunct}
{\mcitedefaultendpunct}{\mcitedefaultseppunct}\relax
\EndOfBibitem
\bibitem{LHCb-PAPER-2020-034}
LHCb collaboration, R.~Aaij {\em et~al.}, \ifthenelse{\boolean{articletitles}}{\emph{{Observation of a new excited $D_s^+$ state in $B^0 \to D^- D^+ K^+\pi^-$ decays}}, }{}\href{https://doi.org/10.1103/PhysRevLett.126.122002}{Phys.\ Rev.\ Lett.\  \textbf{126} (2021) 122002}, \href{http://arxiv.org/abs/2011.09112}{{\normalfont\ttfamily arXiv:2011.09112}}\relax
\mciteBstWouldAddEndPuncttrue
\mciteSetBstMidEndSepPunct{\mcitedefaultmidpunct}
{\mcitedefaultendpunct}{\mcitedefaultseppunct}\relax
\EndOfBibitem
\bibitem{Xie:2021Ds2590}
J.-M. Xie, M.-Z. Liu, and L.-S. Geng, \ifthenelse{\boolean{articletitles}}{\emph{{${D}_{s0}(2590)$ as a dominant $c\overline{s}$ state with a small ${D}^{*}K$ component}}, }{}\href{https://doi.org/10.1103/PhysRevD.104.094051}{Phys.\ Rev.\  \textbf{D104} (2021) 094051}\relax
\mciteBstWouldAddEndPuncttrue
\mciteSetBstMidEndSepPunct{\mcitedefaultmidpunct}
{\mcitedefaultendpunct}{\mcitedefaultseppunct}\relax
\EndOfBibitem
\bibitem{Hao:2022CoupledChannel}
W.~Hao, Y.~Lu, and B.-S. Zou, \ifthenelse{\boolean{articletitles}}{\emph{{Coupled channel effects for the charmed-strange mesons}}, }{}\href{https://doi.org/10.1103/PhysRevD.106.074014}{Phys.\ Rev.\  \textbf{D106} (2022) 074014}\relax
\mciteBstWouldAddEndPuncttrue
\mciteSetBstMidEndSepPunct{\mcitedefaultmidpunct}
{\mcitedefaultendpunct}{\mcitedefaultseppunct}\relax
\EndOfBibitem
\bibitem{Ortega:2022Ds2590}
P.~G. Ortega, J.~Segovia, D.~R. Entem, and F.~Fernandez, \ifthenelse{\boolean{articletitles}}{\emph{{The $D_{s0}(2590)^+$ as the dressed $c\bar{s}(2\,{}^{1}S_{0})$ meson in a coupled-channels calculation}}, }{}\href{https://doi.org/10.1016/j.physletb.2022.136998}{Phys.\ Lett.\  \textbf{B827} (2022) 136998}\relax
\mciteBstWouldAddEndPuncttrue
\mciteSetBstMidEndSepPunct{\mcitedefaultmidpunct}
{\mcitedefaultendpunct}{\mcitedefaultseppunct}\relax
\EndOfBibitem
\bibitem{Yang:2022DsCoupled}
Z.~Yang {\em et~al.}, \ifthenelse{\boolean{articletitles}}{\emph{{Novel Coupled Channel Framework Connecting the Quark Model and Lattice QCD for the Near-threshold ${D}_{s}$ States}}, }{}\href{https://doi.org/10.1103/PhysRevLett.128.112001}{Phys.\ Rev.\ Lett.\  \textbf{128} (2022) 112001}\relax
\mciteBstWouldAddEndPuncttrue
\mciteSetBstMidEndSepPunct{\mcitedefaultmidpunct}
{\mcitedefaultendpunct}{\mcitedefaultseppunct}\relax
\EndOfBibitem
\bibitem{Gao:2022Ds2590}
Z.~Gao {\em et~al.}, \ifthenelse{\boolean{articletitles}}{\emph{{Canonical interpretation of the ${D}_{s0}(2590)^{+}$ resonance}}, }{}\href{https://doi.org/10.1103/PhysRevD.105.074037}{Phys.\ Rev.\  \textbf{D105} (2022) 074037}\relax
\mciteBstWouldAddEndPuncttrue
\mciteSetBstMidEndSepPunct{\mcitedefaultmidpunct}
{\mcitedefaultendpunct}{\mcitedefaultseppunct}\relax
\EndOfBibitem
\bibitem{Yang:2023CharmStrange}
J.~J. Yang, X.~Wang, D.~M. Li {\em et~al.}, \ifthenelse{\boolean{articletitles}}{\emph{{The mass spectrum and strong decay properties of the charmed-strange mesons within Godfrey--Isgur model considering the coupled-channel effects}}, }{}\href{https://doi.org/10.1140/epjc/s10052-023-12275-3}{Eur.\ Phys.\ J.\  \textbf{C83} (2023) 1098}\relax
\mciteBstWouldAddEndPuncttrue
\mciteSetBstMidEndSepPunct{\mcitedefaultmidpunct}
{\mcitedefaultendpunct}{\mcitedefaultseppunct}\relax
\EndOfBibitem
\bibitem{Arifi:2022Mixing}
A.~J. Arifi, H.-M. Choi, C.-R. Ji, and Y.~Oh, \ifthenelse{\boolean{articletitles}}{\emph{{Mixing effects on $1S$ and $2S$ state heavy mesons in the light-front quark model}}, }{}\href{https://doi.org/10.1103/PhysRevD.106.014009}{Phys.\ Rev.\  \textbf{D106} (2022) 014009}\relax
\mciteBstWouldAddEndPuncttrue
\mciteSetBstMidEndSepPunct{\mcitedefaultmidpunct}
{\mcitedefaultendpunct}{\mcitedefaultseppunct}\relax
\EndOfBibitem
\bibitem{Moir2013LQCD}
G.~Moir {\em et~al.}, \ifthenelse{\boolean{articletitles}}{\emph{{Excited spectroscopy of charmed mesons from lattice QCD}}, }{}\href{https://doi.org/10.1007/JHEP05(2013)021}{JHEP \textbf{05} (2013) 021}, \href{http://arxiv.org/abs/1301.7670}{{\normalfont\ttfamily arXiv:1301.7670}}\relax
\mciteBstWouldAddEndPuncttrue
\mciteSetBstMidEndSepPunct{\mcitedefaultmidpunct}
{\mcitedefaultendpunct}{\mcitedefaultseppunct}\relax
\EndOfBibitem
\bibitem{BaBar2006DsJinB}
BaBar collaboration, B.~Aubert {\em et~al.}, \ifthenelse{\boolean{articletitles}}{\emph{{A study of the $D^*_{sJ}(2317)$ and $D_{sJ}(2460)$ mesons in inclusive $c\bar{c}$ production near $\sqrt{s} = 10.6$\,GeV}}, }{}\href{https://doi.org/10.1103/PhysRevD.74.032007}{Phys.\ Rev.\  \textbf{D74} (2006) 032007}, \href{http://arxiv.org/abs/hep-ex/0604030}{{\normalfont\ttfamily arXiv:hep-ex/0604030}}\relax
\mciteBstWouldAddEndPuncttrue
\mciteSetBstMidEndSepPunct{\mcitedefaultmidpunct}
{\mcitedefaultendpunct}{\mcitedefaultseppunct}\relax
\EndOfBibitem
\bibitem{LHCb-DP-2008-001}
LHCb collaboration, A.~A. Alves~Jr.\ {\em et~al.}, \ifthenelse{\boolean{articletitles}}{\emph{{The \lhcb detector at the LHC}}, }{}\href{https://doi.org/10.1088/1748-0221/3/08/S08005}{JINST \textbf{3} (2008) S08005} LHCb-DP-2008-001\relax
\mciteBstWouldAddEndPuncttrue
\mciteSetBstMidEndSepPunct{\mcitedefaultmidpunct}
{\mcitedefaultendpunct}{\mcitedefaultseppunct}\relax
\EndOfBibitem
\bibitem{LHCb-DP-2014-002}
LHCb collaboration, R.~Aaij {\em et~al.}, \ifthenelse{\boolean{articletitles}}{\emph{{LHCb detector performance}}, }{}\href{https://doi.org/10.1142/S0217751X15300227}{Int.\ J.\ Mod.\ Phys.\  \textbf{A30} (2015) 1530022}, \href{http://arxiv.org/abs/1412.6352}{{\normalfont\ttfamily arXiv:1412.6352}}\relax
\mciteBstWouldAddEndPuncttrue
\mciteSetBstMidEndSepPunct{\mcitedefaultmidpunct}
{\mcitedefaultendpunct}{\mcitedefaultseppunct}\relax
\EndOfBibitem
\bibitem{LHCb-DP-2012-004}
R.~Aaij {\em et~al.}, \ifthenelse{\boolean{articletitles}}{\emph{{The \lhcb trigger and its performance in 2011}}, }{}\href{https://doi.org/10.1088/1748-0221/8/04/P04022}{JINST \textbf{8} (2013) P04022}, \href{http://arxiv.org/abs/1211.3055}{{\normalfont\ttfamily arXiv:1211.3055}}\relax
\mciteBstWouldAddEndPuncttrue
\mciteSetBstMidEndSepPunct{\mcitedefaultmidpunct}
{\mcitedefaultendpunct}{\mcitedefaultseppunct}\relax
\EndOfBibitem
\bibitem{Stripping}
N.~Grieser {\em et~al.}, \ifthenelse{\boolean{articletitles}}{\emph{{The LHCb stripping project: Sustainable legacy data processing for high-energy physics}}, }{}\href{https://doi.org/10.1007/s41781-025-00151-6}{Comput.\ Softw.\ Big.\ Sci.\  \textbf{9} (2025) 21}, \href{http://arxiv.org/abs/2509.05294}{{\normalfont\ttfamily arXiv:2509.05294}}\relax
\mciteBstWouldAddEndPuncttrue
\mciteSetBstMidEndSepPunct{\mcitedefaultmidpunct}
{\mcitedefaultendpunct}{\mcitedefaultseppunct}\relax
\EndOfBibitem
\bibitem{LHCb-PAPER-2012-048}
LHCb collaboration, R.~Aaij {\em et~al.}, \ifthenelse{\boolean{articletitles}}{\emph{{Measurements of the \Lb, \Xibm, and \Omegab baryon masses}}, }{}\href{https://doi.org/10.1103/PhysRevLett.110.182001}{Phys.\ Rev.\ Lett.\  \textbf{110} (2013) 182001}, \href{http://arxiv.org/abs/1302.1072}{{\normalfont\ttfamily arXiv:1302.1072}}\relax
\mciteBstWouldAddEndPuncttrue
\mciteSetBstMidEndSepPunct{\mcitedefaultmidpunct}
{\mcitedefaultendpunct}{\mcitedefaultseppunct}\relax
\EndOfBibitem
\bibitem{LHCb-PAPER-2013-011}
LHCb collaboration, R.~Aaij {\em et~al.}, \ifthenelse{\boolean{articletitles}}{\emph{{Precision measurement of \D meson mass differences}}, }{}\href{https://doi.org/10.1007/JHEP06(2013)065}{JHEP \textbf{06} (2013) 065}, \href{http://arxiv.org/abs/1304.6865}{{\normalfont\ttfamily arXiv:1304.6865}}\relax
\mciteBstWouldAddEndPuncttrue
\mciteSetBstMidEndSepPunct{\mcitedefaultmidpunct}
{\mcitedefaultendpunct}{\mcitedefaultseppunct}\relax
\EndOfBibitem
\bibitem{Sjostrand:2007gs}
T.~Sj\"{o}strand, S.~Mrenna, and P.~Skands, \ifthenelse{\boolean{articletitles}}{\emph{{A brief introduction to PYTHIA 8.1}}, }{}\href{https://doi.org/10.1016/j.cpc.2008.01.036}{Comput.\ Phys.\ Commun.\  \textbf{178} (2008) 852}, \href{http://arxiv.org/abs/0710.3820}{{\normalfont\ttfamily arXiv:0710.3820}}\relax
\mciteBstWouldAddEndPuncttrue
\mciteSetBstMidEndSepPunct{\mcitedefaultmidpunct}
{\mcitedefaultendpunct}{\mcitedefaultseppunct}\relax
\EndOfBibitem
\bibitem{Lange:2001uf}
D.~J. Lange, \ifthenelse{\boolean{articletitles}}{\emph{{The EvtGen particle decay simulation package}}, }{}\href{https://doi.org/10.1016/S0168-9002(01)00089-4}{Nucl.\ Instrum.\ Meth.\  \textbf{A462} (2001) 152}\relax
\mciteBstWouldAddEndPuncttrue
\mciteSetBstMidEndSepPunct{\mcitedefaultmidpunct}
{\mcitedefaultendpunct}{\mcitedefaultseppunct}\relax
\EndOfBibitem
\bibitem{Golonka:2005pn}
P.~Golonka and Z.~Was, \ifthenelse{\boolean{articletitles}}{\emph{{PHOTOS Monte Carlo: A precision tool for QED corrections in $Z$ and $W$ decays}}, }{}\href{https://doi.org/10.1140/epjc/s2005-02396-4}{Eur.\ Phys.\ J.\  \textbf{C45} (2006) 97}, \href{http://arxiv.org/abs/hep-ph/0506026}{{\normalfont\ttfamily arXiv:hep-ph/0506026}}\relax
\mciteBstWouldAddEndPuncttrue
\mciteSetBstMidEndSepPunct{\mcitedefaultmidpunct}
{\mcitedefaultendpunct}{\mcitedefaultseppunct}\relax
\EndOfBibitem
\bibitem{Allison:2006ve}
Geant4 collaboration, J.~Allison {\em et~al.}, \ifthenelse{\boolean{articletitles}}{\emph{{Geant4 developments and applications}}, }{}\href{https://doi.org/10.1109/TNS.2006.869826}{IEEE Trans.\ Nucl.\ Sci.\  \textbf{53} (2006) 270}\relax
\mciteBstWouldAddEndPuncttrue
\mciteSetBstMidEndSepPunct{\mcitedefaultmidpunct}
{\mcitedefaultendpunct}{\mcitedefaultseppunct}\relax
\EndOfBibitem
\bibitem{Agostinelli:2002hh}
Geant4 collaboration, S.~Agostinelli {\em et~al.}, \ifthenelse{\boolean{articletitles}}{\emph{{Geant4: A simulation toolkit}}, }{}\href{https://doi.org/10.1016/S0168-9002(03)01368-8}{Nucl.\ Instrum.\ Meth.\  \textbf{A506} (2003) 250}\relax
\mciteBstWouldAddEndPuncttrue
\mciteSetBstMidEndSepPunct{\mcitedefaultmidpunct}
{\mcitedefaultendpunct}{\mcitedefaultseppunct}\relax
\EndOfBibitem
\bibitem{LHCb-PROC-2011-006}
M.~Clemencic {\em et~al.}, \ifthenelse{\boolean{articletitles}}{\emph{{The \lhcb simulation application, Gauss: Design, evolution and experience}}, }{}\href{https://doi.org/10.1088/1742-6596/331/3/032023}{J.\ Phys.\ Conf.\ Ser.\  \textbf{331} (2011) 032023}\relax
\mciteBstWouldAddEndPuncttrue
\mciteSetBstMidEndSepPunct{\mcitedefaultmidpunct}
{\mcitedefaultendpunct}{\mcitedefaultseppunct}\relax
\EndOfBibitem
\bibitem{LHCb-DP-2018-001}
R.~Aaij {\em et~al.}, \ifthenelse{\boolean{articletitles}}{\emph{{Selection and processing of calibration samples to measure the particle identification performance of the LHCb experiment in Run 2}}, }{}\href{https://doi.org/10.1140/epjti/s40485-019-0050-z}{Eur.\ Phys.\ J.\ Tech.\ Instr.\  \textbf{6} (2019) 1}, \href{http://arxiv.org/abs/1803.00824}{{\normalfont\ttfamily arXiv:1803.00824}}\relax
\mciteBstWouldAddEndPuncttrue
\mciteSetBstMidEndSepPunct{\mcitedefaultmidpunct}
{\mcitedefaultendpunct}{\mcitedefaultseppunct}\relax
\EndOfBibitem
\bibitem{PDG2024}
Particle Data Group, S.~Navas {\em et~al.}, \ifthenelse{\boolean{articletitles}}{\emph{{\href{http://pdg.lbl.gov/}{Review of particle physics}}}, }{}\href{https://doi.org/10.1103/PhysRevD.110.030001}{Phys.\ Rev.\  \textbf{D110} (2024) 030001}\relax
\mciteBstWouldAddEndPuncttrue
\mciteSetBstMidEndSepPunct{\mcitedefaultmidpunct}
{\mcitedefaultendpunct}{\mcitedefaultseppunct}\relax
\EndOfBibitem
\bibitem{GBDT}
J.~H. Friedman, \ifthenelse{\boolean{articletitles}}{\emph{Greedy function approximation: a gradient boosting machine}, }{}Ann.\ of Statist.\  (2001) 1189\relax
\mciteBstWouldAddEndPuncttrue
\mciteSetBstMidEndSepPunct{\mcitedefaultmidpunct}
{\mcitedefaultendpunct}{\mcitedefaultseppunct}\relax
\EndOfBibitem
\bibitem{Breiman}
L.~Breiman, J.~H. Friedman, R.~A. Olshen, and C.~J. Stone, {\em Classification and regression trees}, Wadsworth international group, Belmont, California, USA, 1984\relax
\mciteBstWouldAddEndPuncttrue
\mciteSetBstMidEndSepPunct{\mcitedefaultmidpunct}
{\mcitedefaultendpunct}{\mcitedefaultseppunct}\relax
\EndOfBibitem
\bibitem{AdaBoost}
Y.~Freund and R.~E. Schapire, \ifthenelse{\boolean{articletitles}}{\emph{A decision-theoretic generalization of on-line learning and an application to boosting}, }{}\href{https://doi.org/10.1006/jcss.1997.1504}{J.\ Comput.\ Syst.\ Sci.\  \textbf{55} (1997) 119}\relax
\mciteBstWouldAddEndPuncttrue
\mciteSetBstMidEndSepPunct{\mcitedefaultmidpunct}
{\mcitedefaultendpunct}{\mcitedefaultseppunct}\relax
\EndOfBibitem
\bibitem{Hocker:2007ht}
H.~Voss, A.~Hoecker, J.~Stelzer, and F.~Tegenfeldt, \ifthenelse{\boolean{articletitles}}{\emph{{TMVA - Toolkit for Multivariate Data Analysis with ROOT}}, }{}\href{https://doi.org/10.22323/1.050.0040}{PoS \textbf{ACAT} (2007) 040}\relax
\mciteBstWouldAddEndPuncttrue
\mciteSetBstMidEndSepPunct{\mcitedefaultmidpunct}
{\mcitedefaultendpunct}{\mcitedefaultseppunct}\relax
\EndOfBibitem
\bibitem{TMVA4}
A.~Hoecker {\em et~al.}, \ifthenelse{\boolean{articletitles}}{\emph{{TMVA 4 - Toolkit for Multivariate Data Analysis with ROOT. Users Guide}}, }{}\href{http://arxiv.org/abs/physics/0703039}{{\normalfont\ttfamily arXiv:physics/0703039}}\relax
\mciteBstWouldAddEndPuncttrue
\mciteSetBstMidEndSepPunct{\mcitedefaultmidpunct}
{\mcitedefaultendpunct}{\mcitedefaultseppunct}\relax
\EndOfBibitem
\bibitem{Skwarnicki:1986xj}
T.~Skwarnicki, {\em {A study of the radiative cascade transitions between the Upsilon-prime and Upsilon resonances}}, PhD thesis, Institute of Nuclear Physics, Krakow, 1986, {\href{http://inspirehep.net/record/230779/}{DESY-F31-86-02}}\relax
\mciteBstWouldAddEndPuncttrue
\mciteSetBstMidEndSepPunct{\mcitedefaultmidpunct}
{\mcitedefaultendpunct}{\mcitedefaultseppunct}\relax
\EndOfBibitem
\bibitem{Pivk:2004ty}
M.~Pivk and F.~R. Le~Diberder, \ifthenelse{\boolean{articletitles}}{\emph{{sPlot: A statistical tool to unfold data distributions}}, }{}\href{https://doi.org/10.1016/j.nima.2005.08.106}{Nucl.\ Instrum.\ Meth.\  \textbf{A555} (2005) 356}, \href{http://arxiv.org/abs/physics/0402083}{{\normalfont\ttfamily arXiv:physics/0402083}}\relax
\mciteBstWouldAddEndPuncttrue
\mciteSetBstMidEndSepPunct{\mcitedefaultmidpunct}
{\mcitedefaultendpunct}{\mcitedefaultseppunct}\relax
\EndOfBibitem
\bibitem{LHCb-PAPER-2014-035}
LHCb collaboration, R.~Aaij {\em et~al.}, \ifthenelse{\boolean{articletitles}}{\emph{{Observation of overlapping spin-$1$ and spin-$3$ $\Dzb\Km$ resonances at mass $2.86$\gevcc}}, }{}\href{https://doi.org/10.1103/PhysRevLett.113.162001}{Phys.\ Rev.\ Lett.\  \textbf{113} (2014) 162001}, \href{http://arxiv.org/abs/1407.7574}{{\normalfont\ttfamily arXiv:1407.7574}}\relax
\mciteBstWouldAddEndPuncttrue
\mciteSetBstMidEndSepPunct{\mcitedefaultmidpunct}
{\mcitedefaultendpunct}{\mcitedefaultseppunct}\relax
\EndOfBibitem
\bibitem{Lindenbaum:1957ec}
S.~J. Lindenbaum and R.~M. Sternheimer, \ifthenelse{\boolean{articletitles}}{\emph{{Isobaric nucleon model for pion production in nucleon-nucleon collisions}}, }{}\href{https://doi.org/10.1103/PhysRev.105.1874}{Phys.\ Rev.\  \textbf{105} (1957) 1874}\relax
\mciteBstWouldAddEndPuncttrue
\mciteSetBstMidEndSepPunct{\mcitedefaultmidpunct}
{\mcitedefaultendpunct}{\mcitedefaultseppunct}\relax
\EndOfBibitem
\bibitem{Jacob:1959at}
M.~Jacob and G.~C. Wick, \ifthenelse{\boolean{articletitles}}{\emph{{On the general theory of collisions for particles with spin}}, }{}\href{https://doi.org/10.1016/0003-4916(59)90051-X}{Annals Phys.\  \textbf{7} (1959) 404}\relax
\mciteBstWouldAddEndPuncttrue
\mciteSetBstMidEndSepPunct{\mcitedefaultmidpunct}
{\mcitedefaultendpunct}{\mcitedefaultseppunct}\relax
\EndOfBibitem
\bibitem{Chung:1993da}
S.~U. Chung, \ifthenelse{\boolean{articletitles}}{\emph{{Helicity coupling amplitudes in tensor formalism}}, }{}\href{https://doi.org/10.1103/PhysRevD.48.1225}{Phys.\ Rev.\  \textbf{D48} (1993) 1225}, Erratum \href{https://doi.org/10.1103/PhysRevD.56.4419}{ibid.\   \textbf{D56} (1997) 4419}\relax
\mciteBstWouldAddEndPuncttrue
\mciteSetBstMidEndSepPunct{\mcitedefaultmidpunct}
{\mcitedefaultendpunct}{\mcitedefaultseppunct}\relax
\EndOfBibitem
\bibitem{LHCb-PAPER-2015-029}
LHCb collaboration, R.~Aaij {\em et~al.}, \ifthenelse{\boolean{articletitles}}{\emph{{Observation of $\jpsi\proton$ resonances consistent with pentaquark states in \mbox{\decay{\Lb}{\jpsi\proton\Km}} decays}}, }{}\href{https://doi.org/10.1103/PhysRevLett.115.072001}{Phys.\ Rev.\ Lett.\  \textbf{115} (2015) 072001}, \href{http://arxiv.org/abs/1507.03414}{{\normalfont\ttfamily arXiv:1507.03414}}\relax
\mciteBstWouldAddEndPuncttrue
\mciteSetBstMidEndSepPunct{\mcitedefaultmidpunct}
{\mcitedefaultendpunct}{\mcitedefaultseppunct}\relax
\EndOfBibitem
\bibitem{Silverman1986}
B.~W. Silverman, {\em Density estimation for statistics and data analysis}, \href{https://doi.org/10.1007/978-1-4899-3324-9}{ Chapman and Hall, London, 1986}\relax
\mciteBstWouldAddEndPuncttrue
\mciteSetBstMidEndSepPunct{\mcitedefaultmidpunct}
{\mcitedefaultendpunct}{\mcitedefaultseppunct}\relax
\EndOfBibitem
\bibitem{lass}
D.~Aston {\em et~al.}, \ifthenelse{\boolean{articletitles}}{\emph{{A study of $K^-\pi^+$ scattering in the reaction $K^- p\to K^- \pi^+ n$ at $11 \gevc$}}, }{}\href{https://doi.org/10.1016/0550-3213(88)90028-4}{Nucl.\ Phys.\  \textbf{B296} (1988) 493}\relax
\mciteBstWouldAddEndPuncttrue
\mciteSetBstMidEndSepPunct{\mcitedefaultmidpunct}
{\mcitedefaultendpunct}{\mcitedefaultseppunct}\relax
\EndOfBibitem
\bibitem{Wilks:1938dza}
S.~S. Wilks, \ifthenelse{\boolean{articletitles}}{\emph{{The large-sample distribution of the likelihood ratio for testing composite hypotheses}}, }{}\href{https://doi.org/10.1214/aoms/1177732360}{Ann.\ Math.\ Stat.\  \textbf{9} (1938) 60}\relax
\mciteBstWouldAddEndPuncttrue
\mciteSetBstMidEndSepPunct{\mcitedefaultmidpunct}
{\mcitedefaultendpunct}{\mcitedefaultseppunct}\relax
\EndOfBibitem
\bibitem{FOCUS:2007mcb}
FOCUS collaboration, J.~M. Link {\em et~al.}, \ifthenelse{\boolean{articletitles}}{\emph{{Dalitz plot analysis of the $D^{+} \to K^{-} \pi^{+} \pi^{+}$ decay in the FOCUS experiment}}, }{}\href{https://doi.org/10.1016/j.physletb.2007.06.070}{Phys.\ Lett.\  \textbf{B653} (2007) 1}, \href{http://arxiv.org/abs/0705.2248}{{\normalfont\ttfamily arXiv:0705.2248}}\relax
\mciteBstWouldAddEndPuncttrue
\mciteSetBstMidEndSepPunct{\mcitedefaultmidpunct}
{\mcitedefaultendpunct}{\mcitedefaultseppunct}\relax
\EndOfBibitem
\end{mcitethebibliography}
